\documentclass[english,prb, twocolumn, superscriptaddress]{revtex4-2}
\usepackage[T1]{fontenc}
\usepackage[utf8]{inputenc}
\usepackage{verbatim}
\usepackage{amsmath}
\usepackage{graphicx}
\usepackage{amssymb}
\usepackage{xcolor}
\usepackage{soul}
\usepackage{bm}
\renewcommand{\mathbf}{\bm}
\makeatletter
\usepackage{hyperref}
\makeatother

\usepackage{babel}
\begin{document}

\title{Stability of moving solitons in trans-polyacetylene in an electric field}

\author{Leandro M. Arancibia}
\affiliation{Instituto Interdisciplinario de Ciencias B\'{a}sicas - Consejo Nacional de Investigaciones Cient\'{i}ficas y T\'{e}cnicas }
\affiliation{Facultad de Ciencias Exactas y Naturales - Universidad Nacional de Cuyo}

\author{Cristi\'{a}n G. S\'{a}nchez}
\affiliation{Instituto Interdisciplinario de Ciencias B\'{a}sicas - Consejo Nacional de Investigaciones Cient\'{i}ficas y T\'{e}cnicas }
\affiliation{Facultad de Ciencias Exactas y Naturales - Universidad Nacional de Cuyo}

\author{Alejandro M. Lobos}
\affiliation{Instituto Interdisciplinario de Ciencias B\'{a}sicas - Consejo Nacional de Investigaciones Cient\'{i}ficas y T\'{e}cnicas }
\affiliation{Facultad de Ciencias Exactas y Naturales - Universidad Nacional de Cuyo}

\date{\today}

\begin{abstract}
In this work we study the dynamics and stability of charged solitons in trans-polyacetylene (tPA), and revisit the issue of the stability of these non-linear excitations under the effect of an external electric field applied parallel to the polymer. Using the formalism of the Su-Schrieffer-Heeger (SSH) model, we solve the coupled dynamical equations for electrons and classical nuclei at the mean-field level and in the regime of low external electric field $E$, where the dynamics of the moving soliton is adiabatic. Analyzing observable quantities in real space and frequency space, we identify the microscopic mechanisms triggering the dynamical instabilities of the soliton. In addition, we put forward the definition of a proper quantitative measure of its stability, an issue which to the best of our knowledge has remained an open question. Besides its intrinsic interest from the fundamental point of view, our work might be relevant for the design of novel organic electronic devices based on soliton-mediated transport.
\end{abstract}


\maketitle

\section{Introduction}
Trans-polyacetylene (tPA), a linear chain of carbon atoms with alternating single and double bonds, has attracted the interest of the scientific community for more than 40 years \citep{heeger1988solitons}. Apart from being the simplest of organic conductors, a  fascinating feature of this system is the presence of solitonic excitations (i.e., self-trapped combinations of localized electronic states and ion-lattice distortions) with fractional charge $e/2$ and spin $S=0$ which originate in its doubly-degenerate ground state. The presence of these solitonic excitations in tPA was indirectly established via optical spectroscopy\citep{sethna1982photoinduced, blanchet1983photoexcitations}, and magnetic (EPR) experiments\citep{goldberg1979electron, weinberger1980electron}. Recently, scanning tunneling microscopy has enabled the \emph{direct} observation of soliton-like structures for the first time on individual tPA molecules synthesized on top a Cu(111) surface \cite{Wang19_Solitons_in_individual_PA_molecules}. This constitutes an important achievement in the experimental study of solitons in condensed-matter physics, with potential applications to organic electronics
\cite{Farchioni_Organic_Electronic_Materials}.


On the theoretical side, the so-called Su-Schrieffer-Heeger (SSH) Hamiltonian \cite{Su79_Solitons_in_polyacetylene}, a simple model encoding the interaction between electronic and classical ionic degrees of freedom, has successfully accounted for many of the experimentally observed properties of tPA, and theoretically explains the emergence of fractional solitons. The SSH model is one of the most paradigmatic models in condensed matter physics as it realizes the simplest example of a topological insultator in one dimension \cite{Bernevig_book_TI_TSC, Asboth16_Short_course_on_TIs}. 
Within this theoretical framework, particular attention has been devoted to the understanding of moving solitons in tPA in out-of-equilibrium  conditions. In this respect, several works\cite{Su80_Dynamics_of_solitons_in_PA, Mele82_Hot_luminiscence_in_PA, Bishop84_Breathers_in_Polyacetylene, Phillpot87_Dynamics_in_polyyne} studied transient phenomena in tPA chains with photogenerated soliton-antisoliton excitations, and pointed to the existence of a maximum velocity $v_\text{max}\simeq 2.7 v_s$, with $v_s$ the velocity of sound along the tPA chain, beyond which the moving soliton solution becomes unstable \cite{Bishop84_Breathers_in_Polyacetylene, Phillpot87_Dynamics_in_polyyne}. 
Later, other authors \cite{ono1990motion, Rakhmanova99_Soliton_dissociation_in_high_electric_field, Johansson04_Nonadiabatic_simulation_of_polaron_dynamics} studied the motion of individual charged solitons on tPA under the effect of an external uniform electric field $E$. An important conclusion in those works is that at high enough electric fields, the electronic degrees of freedom decouple from the lattice distortions, rendering the soliton unstable. In particular, based on their numerical results, Ono and Terai\cite{ono1990motion} concluded that the center-of-mass velocity of the moving soliton reaches a saturation value $v_\text{sat}\simeq 4 v_s$ (slightly above the previous prediction), which is \textit{independent} of the applied electric field. This result has remained as an empirical upper limit for the soliton velocity in the presence of an electric field $E$. However, an intuitive argument suggests that the saturation velocity $v_\text{sat}$ should drop to zero in the limit of vanishing field $E\rightarrow 0$ in order to continuously match with the equilibrium condition. Therefore, a more careful study in this regime is desirable.

In addition to its intrinsic interest for fundamental physics, understanding soliton-mediated transport might have important implications for technological applications in organic electronic devices (photovoltaic cells, FETs, LEDs, etc). 
However, the non-linear dynamics of moving solitons represents a major theoretical challenge due to its inherent complexity and the large number of interacting degrees of freedom. Some of the open questions in this context are: How to quantitatively determine the stability of dynamical solitons in an out-of-equilibrium situation? What are the main mechanisms driving the instability?

Motivated by these recent experimental developments and by these open questions, in this work we revisit the issue of the stability of solitons in tPA under non-equilibrium conditions. In particular, we focus on the effect of an external electric field applied parallel to the tPA chain, and on the experimentally relevant question of  the maximum electric field (instead of the maximal velocity) that a solitonic excitation can support in ideal conditions. We solve the coupled dynamical equations for electrons and classical nuclei at the mean-field (i.e., Ehrenfest approximation) level, and study the motion of an externally driven charged soliton. We focus in the regime of low external electric field $E$, where the dynamics of the moving soliton is adiabatic and electronic interband transitions are frozen. In addition, in order to simplify the numerical calculations, we assume the regime of extremely low doping, where essentially the tPA hosts only a single soliton excitation. By comparing the time-dependent lattice defomation field, and the instantaneous energy spectrum of in-gap electronic excitations, we conclude that the main destabilizing mechanism in the adiabatic regime is the proliferation of soliton-antisoliton pairs along the chain, originated in large-amplitude oscillations of the lattice deformation field (i.e., dynamically excited phonon modes). We put forward a quantitative indicator of soliton stability, suitable for a dynamical situation, and obtain a criterion of stability as a function of the applied longitudinal field $E$. This result might be important for the design of polymer-based electronic circuits. 

We stress that our assumptions are conceptually different from previous works where the moving soliton was assumed to travel at a constant velocity $v$ \textit{in the absence of external fields} (see e.g., Refs.\onlinecite{ Bishop84_Breathers_in_Polyacetylene, Takayama80_Continuum_model_for_PA, Guinea84_Dynamics_of_PA_chains,ye1992vibrational}). Finally, our work resolves inconsistencies in previous works regarding the saturation velocity at very low fields, where we recover the physical result $v_\text{sat}\rightarrow 0$  in the limit $E\rightarrow 0$. 

This article is organized as follows: In Sec. \ref{sec:theory} we present the theoretical framework of the SSH Hamiltonian and the theoretical tools used in this work. In Sec. \ref{sec:numerical} we give technical details of the numerical methods tools used in the integration of the equations of motion. In Sec. \ref{sec:results} we present the main results, and in Sec. \ref{sec:finite_size} we study the finite-size effects on the stability and dynamics of the mobile soliton. Finally, in Sec. \ref{sec:summary} we give our conclusions and perspectives.

\section{Theoretical Framework}\label{sec:theory}
We model an isolated finite tPA chain with $N$ $\text{-(CH)}$ (carbon and hydrogen) groups by means of the SSH model\citep{Su79_Solitons_in_polyacetylene,ono1990motion}  
\begin{align}\label{eq:hamiltonian}
H_{\textup{SSH}}(\phi) &= -\sum_s\sum_{n=1}^N \biggl\{ \left[ t_0 - \alpha ( u_{n+1}-u_n) \right] e^{-i\phi} c_{n+1,s}^\dagger c_{n,s} \nonumber \\&+ \text{H.c.} \biggr\}+\frac{K}{2}\sum_{n=1}^N (u_{n+1}-u_n)^2+ \frac{M}{2}\sum_{n=1}^N \dot{u}_n^2.
\end{align}
Here, the effective site $n$ represents the $n$-th group $\text{-(CH)}_n$ in the tPA chain. The classical variable $u_n$ is the displacement of the $\text{-(CH)}_n$ group along the axis of the polymer from its equilibrium position, and $c_{n,s}^\dagger (c_{n,s})$ is the creation (annihilation) operator for an electron in the $2p_z$ orbital at the C atom on site $n$, with spin projection $s=\uparrow,\downarrow$ along the $z$-axis. The term in brackets in the first line corresponds to the hopping transference integral $t_{n+1,n}$, expanded to first order in terms of the small parameter $(u_{n+1}-u_n)$, $t_0$ is the zeroth-order hopping parameter, and $\alpha$ is the electron-phonon coupling. The phase factor $e^{i\phi(t)}$ appears due to the Peierls substitution, where $\phi(t)=eaA(t)/\hbar c$ depends on the vector potential $A(t)$, whose time derivative is the electric field $E=-\dot{A}/c$. Here $a$ is the lattice parameter in the tPA chain. The second line in Eq. (\ref{eq:hamiltonian}) corresponds to the potential and kinetic energy of -(CH) groups, respectively, where $K$ is the effective spring force constant of the C-C bonds, and $M$ is their mass. To simplify the calculations, in what follows we assume the system at zero temperature, and periodic boundary conditions $u_{N+1}=u_1$, and $c_{N+1,s}=c_{1,s}$, are imposed.

We use standard values generally accepted for experimental tPA systems\cite{ono1990motion}: $t_0=2.5$ eV, $K=21\ \text{eV}/$\AA$^2$, $\alpha=4.1\ \text{eV}/$\AA, and $a=1.22$ \AA. Then, the bare optical phonon frequency $\omega_0=\sqrt{4 K/M}$ is equal to $2.5\times 10^{14}\ s^{-1}$, and the sound velocity of acoustic phonons $v_s=\omega_0 a/2$ equals $1.53\times 10^6$ cm/s. In what follows, we use the parameters $a$ and $\omega_0^{-1}$ as the units of length and time, respectively. In addition, the parameter $E_0=\hbar \omega_0/e a$ is the unit of electric field estimated for the present case as $1.3 \times 10^7$ V/cm.

In order to obtain a physically meaningful initial condition, we focus on the equilibrium properties of the system and we therefore set $\phi=0$ (i.e., absence of external field). 
To that end, we define the \textit{lattice deformation field} for a given configuration of the -(CH) groups as the vector of distances  between neighboring sites $y_n\equiv u_{n+1}-u_n$. In addition, denoting the many-particle ground state as  $|\Psi_\mathrm{gs}\rangle$ and the ground-state energy as $E_\mathrm{gs}$, the stability condition given by the Hellmann-Feynman theorem becomes\citep{feynman1939forces}
\begin{align}
\frac{ \partial E_\mathrm{gs}}{\partial y_n}&=\langle \Psi_\mathrm{gs}| \frac{ \partial H_{\textup{SSH}}}{\partial y_n} |\Psi_\mathrm{gs}\rangle =0.
\end{align}
This expression results in the self-consistent equation
\begin{align}
y_n=&-2\frac{\alpha}{K}\sum_{s}\sum_{\nu=1}^\textup{occ} \left[\psi_{\nu,s}(n)\psi_{\nu,s}(n+1) \right.\nonumber \\
& - \left.\frac{1}{N}\sum_{n=1}^N \psi_{\nu,s}(n)\psi_{\nu,s}(n+1)\right.]\label{eq:self_consistent},
\end{align}
where "occ" stands for occupied states, which correspond to $N_\text{el}$ electrons, and $\psi_{\nu,s}\left(n\right)$ is the amplitude of the $\nu$-th eigenstate $|\psi_{\nu,s}\rangle \equiv \sum_{n=1}^N \psi_{\nu,s}\left(n\right) c^\dagger_{n,s}|0\rangle$ satisfying the eigenvalue equation
\begin{align}\label{eq:eigenvalue_equation}
H_\text{SSH}|\psi_{\nu,s}\rangle&=E_{\nu,s}|\psi_{\nu,s}\rangle.
\end{align}
Note that this equation is solved for a given specific configuration of the lattice deformation field ${y_n}$. We numerically solve Eqs. (\ref{eq:self_consistent}) and (\ref{eq:eigenvalue_equation}) by an iterative method until convergence of the lattice deformation field, obtained when the modulus of the vector $\left(\Delta y\right)_n \equiv y^{(k)}_n - y^{(k-1)}_n$, with $k$ the iteration index, is $\left\Vert \Delta y \right\Vert <10^{-9} a$. From the solution of these equations, the groundstate configuration of the complete system ($N_\text{el}$ electrons plus $N$ ions) is obtained.

Interestingly, when the system is filled with $N_\text{el}$ electrons equal to the number of sites $N$ (i.e., neutral system at half-filling of the conduction band), the self-consistent  solution of Eq. (\ref{eq:self_consistent}) reproduces the well-known Peierls instability  with a stable dimerized lattice configuration given by the expression $y_n=\left(-1\right)^n y_0$ \cite{Peierls_Quantum_Theory_of_Solids}, with $y_0=0.06a$. Concomitantly, a single-particle gap of size $2\Delta_g=4\alpha y_0=1.31$ {eV} opens at the Fermi energy, and the system becomes an insulator. At this point, we note that periodic boundary conditions have important consequences in the nature of the ground state of the system: whereas for an even $N$ the aforementioned length alternation of the C-C bonds is compatible with the periodic boundary conditions, for odd $N$ this is not possible and a domain wall in the lattice deformation field $y_n$ naturally emerges. In this last case, a mid-gap zero-energy state localized at the domain-wall appears in the single-particle spectrum \citep{heeger1988solitons, Su79_Solitons_in_polyacetylene}. The presence of this state can also be understood due to the chiral (or sub-lattice) symmetry of the SSH model in the dimerized phase, which generates a symmetric spectrum around the Fermi energy: since for every finite-energy state at $E_{\nu,s}$, another state with energy $-E_{\nu,s}$ must exist, in the presence of an odd number of sites $N$, a state with $E_{\nu,s}=0$ must emerge\cite{Bernevig_book_TI_TSC, Asboth16_Short_course_on_TIs}.

We now turn to the dynamics of the system under the action of an external electric field $E$, which we assume to be suddenly turned on at $t=0$. The reason why this is compatible with an adiabatic time-evolution of the moving soliton is that the Hamiltonian (\ref{eq:hamiltonian}) does not depend on the bare field $E(t)$, but on the vector potential $A(t)$, which is an integrated quantity (a more detailed discussion on the adiabatic approximation is given below). The dynamical equations for electronic and lattice degrees of freedom are solved self-consistently within the Ehrenfest approach\cite{ono1990motion, todorov2001time}. We introduce the usual Born-Oppenheimer approximation, in which the electrons ``move'' according to the instantaneous potential generated by the ions. Additionally, we treat the ions as classical particles obeying Newton's law. The effect of the electrons appears in the form of a ``quantum force'', obtained by solving the Hamilton-Jacobi equations in the mean-field approximation: 
\begin{align}\label{eq:hamilton_jacobi}
\dot{p}_n\left(t\right) &=- \langle \Psi_\text{e}\left(t\right)| \frac{ \partial H_{\textup{SSH}}}{\partial u_n} |\Psi_\text{e}\left(t\right)\rangle,\\\nonumber
\dot{u}_n\left(t\right)&=p_n\left(t\right)/M.
\end{align}
Here  $|\Psi_\text{e}\left(t\right)\rangle$ is the many-electron state vector obeying the time-dependent Schr\"odinger equation 
\begin{equation}\label{eq:tdschroedinger}
i\hbar |\dot{\Psi}_\text{e}\left(t\right)\rangle =H_{\textup{SSH}}\left(\{y_n(t)\},t\right)|\Psi_\text{e}\left(t\right)\rangle.
\end{equation}
Alternatively, introducing the time-evolution operator  ${U}(t,t_0)$ which satisfies the equation
\begin{equation}\label{eq:evolution_operator}
i\hbar \dot{U}(t,0)=H_{\textup{SSH}}\left(\{y_n(t)\},t\right)U(t,0),
\end{equation}
and whose formal solution is
\begin{equation}\label{eq:evolution_operator_integral}
U\left(t,0\right) =T e^{-\frac{i}{\hbar}\int_0^t d\tau \ H_{\textup{SSH}}\left(\{y_n(\tau)\},\tau\right)},
\end{equation}
with $T$ the time-ordering operator, the evolved many-particle state can be expressed as 
$|\Psi_\text{e}\left(t\right)\rangle =U\left(t,0\right)|\Psi_\text{gs}\rangle$.

Solving Eqs. (\ref{eq:hamilton_jacobi}) and transforming from $\{u_n\}$ to the $\{y_n\}$ basis, we obtain \cite{ono1990motion}
\begin{equation}\label{eq:newton}
\begin{split}
\ddot{y}_n(t)=& -\frac{K}{M}(2y_n(t)-y_{n-1}(t)-y_{n+1}(t)) \\
& + \frac{\alpha}{M}\sum_{s}\sum_{\nu}^\textup{occ}e^{i\phi}\left\lbrace \psi^*_{\nu,s}(t,n+1)\psi_{\nu,s}(t,n+2) -\right.\\
&\left. 2 \psi^*_{\nu,s}(t,n)\psi_{\nu,s}(t,n+1) +\psi^*_{\nu,s}(t,n-1)\psi_{\nu,s}(t,n)\right\rbrace \\
& + \text{h.c.}	
\end{split}
\end{equation}
Note that Eq. (\ref{eq:newton}) reduces to Eq. (\ref{eq:self_consistent}) in the static case and in the absence of external fields. Eqs. (\ref{eq:evolution_operator}) and (\ref{eq:newton}), with the solution of Eq. (\ref{eq:self_consistent}) as the initial condition at time $t=0$, form a coupled set of nonlinear differential equations. This theoretical approach is usually known as ``Ehrenfest dynamics''.

\subsection{Moving soliton in the adiabatic approximation}\label{sec:adiabatic_evolution}


In addition to the above considerations, in this work we will focus on the regime of low field $E$. This enables to introduce the adiabatic approximation\cite{messiah}. Under the hypothesis of sufficiently slow changes of the Hamiltonian from $H_{\text{SSH}}(0)$ to $H_{\text{SSH}}(2\pi)$, and in the case of a discrete and non-degenerate spectrum, the adiabatic approximation provides a way to describe the time-dependent wavefunctions in terms of the instantaneous eigenfunctions of $H_{\text{SSH}}(\phi)$. In particular, the independent single-particle states evolve according to the expression given by the adiabatic theorem \cite{messiah}:
\begin{align}
\label{eq:adiabatic_theorem}
\left|\psi_{\nu,s}\left(t\right)\right\rangle  & \simeq e^{i\theta_{\nu}\left(t\right)}e^{i\gamma_{\nu}\left(t\right)}\left|\phi_{\nu,s}\left(t\right)\right\rangle ,
\end{align}
where the states $\left|\phi_{\nu,s}\left(t\right)\right\rangle$ are the instantaneous eigenstates of  $H_{\text{SSH}}(t)$, i.e. $H_{\text{SSH}}(t) \left|\phi_{\nu,s}\left(t\right)\right\rangle = \epsilon_\nu(t)\left|\phi_{\nu,s}\left(t\right)\right\rangle$, 
where the time appears as a parameter, and
\begin{align*}
\theta_{\nu}\left(t\right) & =-\frac{1}{\hbar}\int_{0}^{t}d\tau\ \epsilon_{\nu}\left(\tau\right),\\
\gamma_{\nu}\left(t\right) & =i\int_{0}^{t}d\tau\ \left\langle \phi_{\nu,s}\left(\tau\right)\right|\left.\frac{\partial\phi_{\nu,s}}{\partial\tau}\left(\tau\right)\right. \rangle, 
\end{align*}
are the dynamical and Berry phases, respectively. For a generic dynamical state vector $|\psi (t)\rangle$ expressed as a linear combination of the instantaneous eigenstates, $|\psi (t)\rangle=\sum_\nu \alpha_\nu(t)|\phi_\nu (t)\rangle$ (we omit spin indices for simplicity), the validity of the adiabatic theorem Eq. (\ref{eq:adiabatic_theorem}) can be ensured as long as the matrix elements
\begin{align}\label{eq:adiabatic_criterion} 
\frac{\left\langle \phi_\nu \left(t\right)|\dot{H}_\text{SSH}\left(t\right)|\phi_{\mu}\left(t\right)\right\rangle }{\hbar \left( \epsilon_{\nu}\left(t\right)-\epsilon_{\mu}\left(t\right)\right)^2} & \ll 1,
\end{align}
are small enough \cite{messiah}. 
An important observation for our case is that this condition is generically violated for bulk states in a system in the thermodynamical limit (i.e., $N\rightarrow \infty$), since the energy difference becomes infinitesimally small $\Delta \epsilon \propto N^{-2}$. However, a key point is that the zero-energy state localized at the domain wall,  which is separated from the bulk states by half the quasiparticle gap $\Delta_g$, does not suffer from this problem. In other words, the energy gap $\Delta_g$ dynamically ``protects''  the zero-energy state. This is not surprising given the topological nature of the zero-energy state in the SSH Hamiltonian\cite{Bernevig_book_TI_TSC, Asboth16_Short_course_on_TIs}. Therefore, particularizing condition (\ref{eq:adiabatic_criterion}) for the zero-energy instantaneous state, we obtain the following order-of-magnitude estimation for the field $E$ in the adiabatic regime (see details in the Appendix):
\begin{align}\label{eq:stability_condition2}
E \ll E_\text{c} = \frac{\Delta^2_g}{4t_0ea} \simeq 0.260 E_0
\end{align}

In this regime a moving soliton will evolve adiabatically, and therefore as a stable quantum object. For the present parameters this is within the regime of interest of electronic devices\citep{Mahani17_Breakdown_of_polarons_in_electric_field}. Note that the above condition is essentially the same condition to avoid the electric breakdown of an insulator, given by the expression \cite{ashcroft}
\begin{align}\label{eq:electric_breakdown}
E&\ll \frac{\Delta^2}{t_0 e a},
\end{align}
with $\Delta=2\Delta_g$. 

\section{Numerical Methods}\label{sec:numerical}

In order to solve the coupled dynamics of the lattice-electron system, we first define a time step $\Delta t$, chosen sufficiently smalls so that all the dynamical observables of interest converge within acceptable tolerance, and discretize the time variable in $N_\text{step}$ steps as $t\rightarrow t_m=m \Delta t$, where $m=0, 1, \dots, N_\text{step}-1$. In particular, to solve the dynamics of the -(CH) groups [Eq. (\ref{eq:newton})] we implemented the Verlet's velocity algorithm 
\begin{align}
y_n(t_{m+1})=& y_n(t_m)+\dot{y}_n(t_m)\Delta t+\frac{1}{2}\ddot{y}_n(t_m)\Delta t^2\\ 
\dot{y}_n(t_{m+1}) =& \dot{y}_n(t_m) + \frac{\ddot{y}_n(t_{m+1})+\ddot{y}_n(t_m)}{2}\Delta t,\label{eq:acceleration}
\end{align}
which allows to obtain the new quantities $y_n(t_{m+1})$ and  
$\dot{y}_n(t_{m+1})$ in terms of the old ones, i.e., $y_n(t_{m})$ and  
$\dot{y}_n(t_{m})$, and the force given in Eq. (\ref{eq:hamilton_jacobi}). As mentioned before, the initial condition for our simulations corresponds to the lattice configuration $y_n(0)$ taken from the static solution of Eq. (\ref{eq:self_consistent}), and $\dot{y}_n(0)=0$ (zero initial velocity of the -(CH) groups).

The time-evolution operator Eq. (\ref{eq:evolution_operator_integral}) can be Taylor-expanded  for sufficiently small $\Delta t$ using a midpoint scheme (i.e., the Crank-Nicolson algorithm\cite{scherer2017computational})  
\begin{align}
U\left(t_{m+1},t_m\right)\simeq & \left( I + \frac{i\Delta t}{2\hbar} H_{\textup{SSH}}\left(t_{m+1/2}\right) \right)^{-1} \nonumber \\
&\times \left( I - \frac{i\Delta t}{2\hbar} H_{\textup{SSH}}\left(t_{m+1/2}\right) \right)
\label{eq:crank_nicolson}
\end{align}
where $I$ is the identity operator, and the Hamiltonian is evaluated at the midpoint $t_{m+1/2}\equiv  t_m+\Delta t/2$ by the means of a simple Euler step integration. After some algebra, a time evolution $\Delta t$ for the electronic wavefunctions can be expressed as 
\begin{align}\label{eq:wavefunction_evolution}
|\psi_{\nu ,s}(t_{m+1})\rangle=& 2 \left( I + \frac{i\Delta t}{2\hbar} H_{\textup{SSH}}(t_{m+1/2}) \right)^{-1}|\psi_{\nu ,s}(t_m)\rangle\\ \nonumber & - |\psi_{\nu ,s}(t_m)\rangle.
\end{align}
The Crank-Nicolson method has the advantage of being very simple to implement, while avoiding the need to diagonalize the spectrum at each time step \cite{scherer2017computational}. 
In practical terms, the actual value of $\Delta t$ used in our simulations emerges from a compromise between the pursuit of numerical convergence of the quantities of interest (which is satisfied for a large $N_\text{step}$), and the numerical cost (which is minimized for a small $N_\text{step}$). We have numerically confirmed that the choice $\Delta t=0.01\ \omega_0^{-1}$ is sufficient to reach convergence in all the observables within a tolerance of 0.2\%.

\section{Results and discussion \label{sec:results}}

\begin{figure*}
\includegraphics[width=0.9\textwidth]{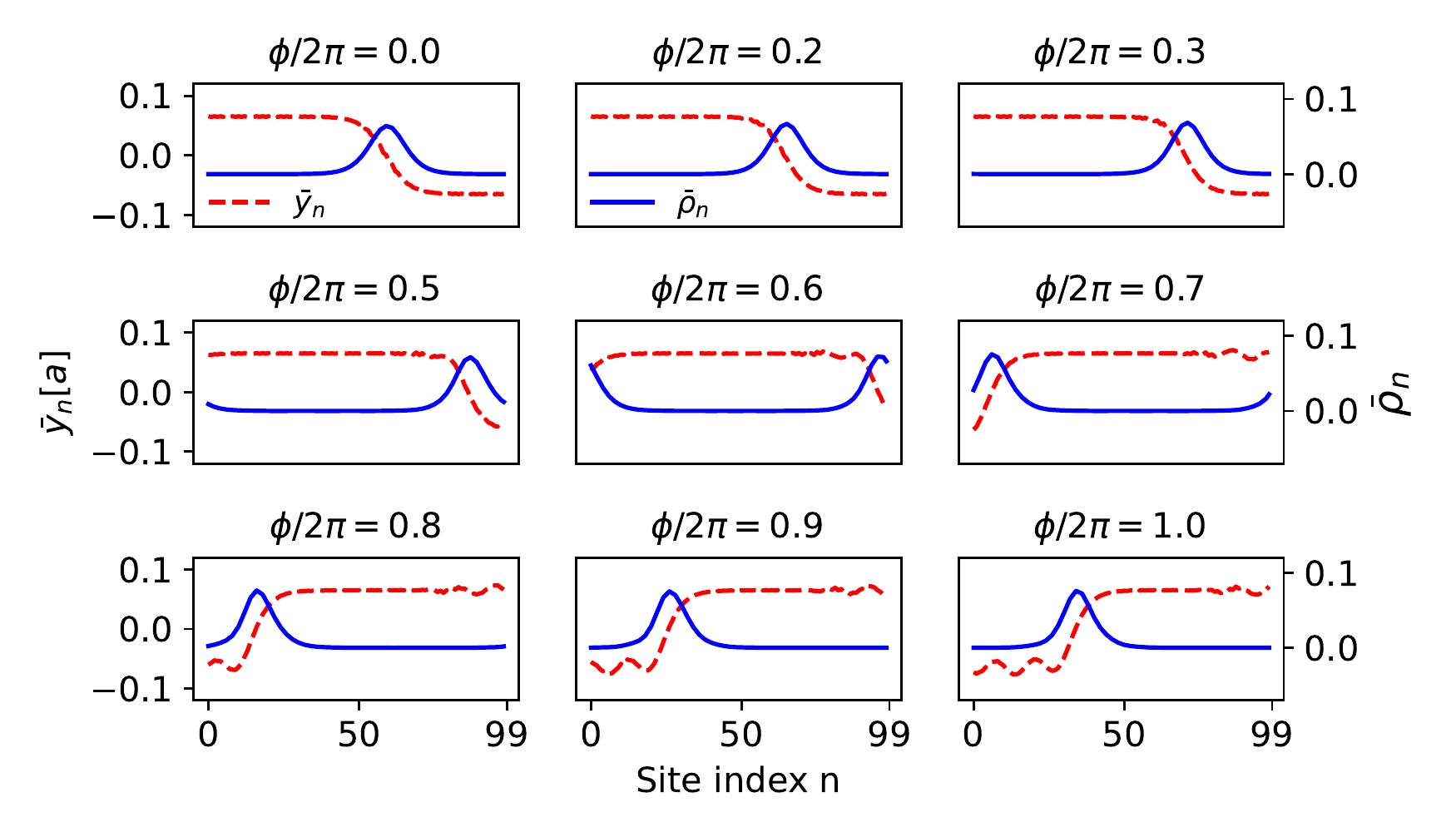}
\caption{Staggered lattice deformation field $\bar{y}_n$ (dashed red line) [Eq. (\ref{eq:staggered_field})] and electronic  excess charge $\bar{\rho}_n$ (continuous blue line) [Eq. (\ref{eq:excess_charge})] 
versus site index $n$ of the tPA chain. An external electric field $E=0.01E_0$ is suddenly turned on at time $t=0$ and both $\bar{y}_n$  and  $\bar{\rho}_n$  are computed at different times (or, equivalently, at different values of the Peierls phase $\phi$ from $0$ to $2 \pi$). At $\phi=0$, $\bar{y}_n(t)$ shows a kink (i.e., domain wall) connecting the two minima given by the Peierls dimerization instability, and the excess charge $\bar{\rho}_n$  localizes precisely at the center of the domain wall. As the system evolves, despite the noticeable development of a periodic structure  (i.e., phonon mode with well-defined wavelength), the domain-wall profile of $\bar{y}_n$ is roughly preserved, and the center of mass of $\bar{\rho}_n$ follows the position of the center of the domain wall. This self-organized soliton structure is stable and moves as a single object until the end of the simulation (Peierls phase $\phi=2 \pi$). \label{fig:espacio_real_E_p01} }
\end{figure*}

Our goal is to solve the equations of motion in the presence of a uniform external field $E$, which is turned on at $t=0$, and to study the dynamics and stability of the moving solitonic excitation. To that end, it is customary to focus on certain specific dynamical quantities, in particular, the ``optical'' (i.e., staggered) lattice configuration, defined as\cite{Su79_Solitons_in_polyacetylene, ono1990motion,Su80_Dynamics_of_solitons_in_PA,Bishop84_Breathers_in_Polyacetylene}
\begin{equation}\label{eq:staggered_field}
\bar{y}_n\left(t\right) \equiv (-1)^ny_n\left(t\right).
\end{equation} 
This quantity allows the identification of the specific alternation pattern at site $n$, and therefore, to identify the location of the domain walls in the system at a particular time $t$. In the static case and in the continuum limit, this quantity corresponds to the well-known expression $\Delta (x) = u_0 \tanh(x/\xi)$ for a domain-wall of width $\xi$ centered at $x=0$ \cite{heeger1988solitons}. The constant values $\pm u_0$ at $x\rightarrow \pm \infty$ respectively correspond to the two different degenerate minima of the energy (i.e., the two possible Peierls dimerizations). We mention in passing that the related formula\cite{Bishop84_Breathers_in_Polyacetylene, Takayama80_Continuum_model_for_PA,Guinea84_Dynamics_of_PA_chains,ye1992vibrational,Kuwabara96_Semi_Phenomenological_Analysis_of_Nonlinear_Excitations_in_PA} 
\begin{equation}\label{eq:moving_soliton}
\Delta\left(x,t\right) = u_0 \tanh\left(\frac{x-vt}{\xi(v)}\right),
\end{equation} 
has been used in previous works as an ansatz to study the stability of a moving kink traveling at a constant velocity $v$ by means of analytical methods. 

In addition, the (smoothened) dynamical  electronic excess charge, defined as
\begin{equation}\label{eq:excess_charge}
\bar{\rho}_n\left(t\right)\equiv \frac{1}{4	}\left[2\rho_n\left(t\right) + \rho_{n+1}\left(t\right)+\rho_{n-1}\left(t\right)\right],
\end{equation}
where the excess charge density $\rho_n\left(t\right) = \sum_s\sum_{\nu=1}^\textup{occ} |\psi_{\nu,s}(t,n)|^2 - 1$ verifies the sum rule $\sum_n \rho_n\left(t\right) = 0$, intuitively provides information about the localization of the extra charge in real space. 


Most of the simulations in this work have been obtained for $N=99$ ions and $N_\text{el}=100$ electrons (doped tPA with excess charge $Q=-e$). 
 Therefore, under these conditions we ensure that our simulations start with a single charged soliton in the system at $t=0$.  In order to compare simulations obtained for different values of the external field $E$ we look at the accumulated Peierls phase $\phi(t)=eaA(t)/\hbar c=- Et\omega_0/E_0 $ in each case, where the fact that the external field is uniform and constant, is used. Analyzing the dynamical state of the system in terms of the Peierls phase (instead of directly looking at the simulation time) provides a more systematic way to compare different physical situations originated from the different values of $E$. In particular, we define a complete evolution cycle in the system for a total accumulated Peierls phase equal to $2 \pi$, and therefore $t_f=2\pi E_0/E \omega_0$.

\begin{figure*}
\includegraphics[width=0.9\textwidth]{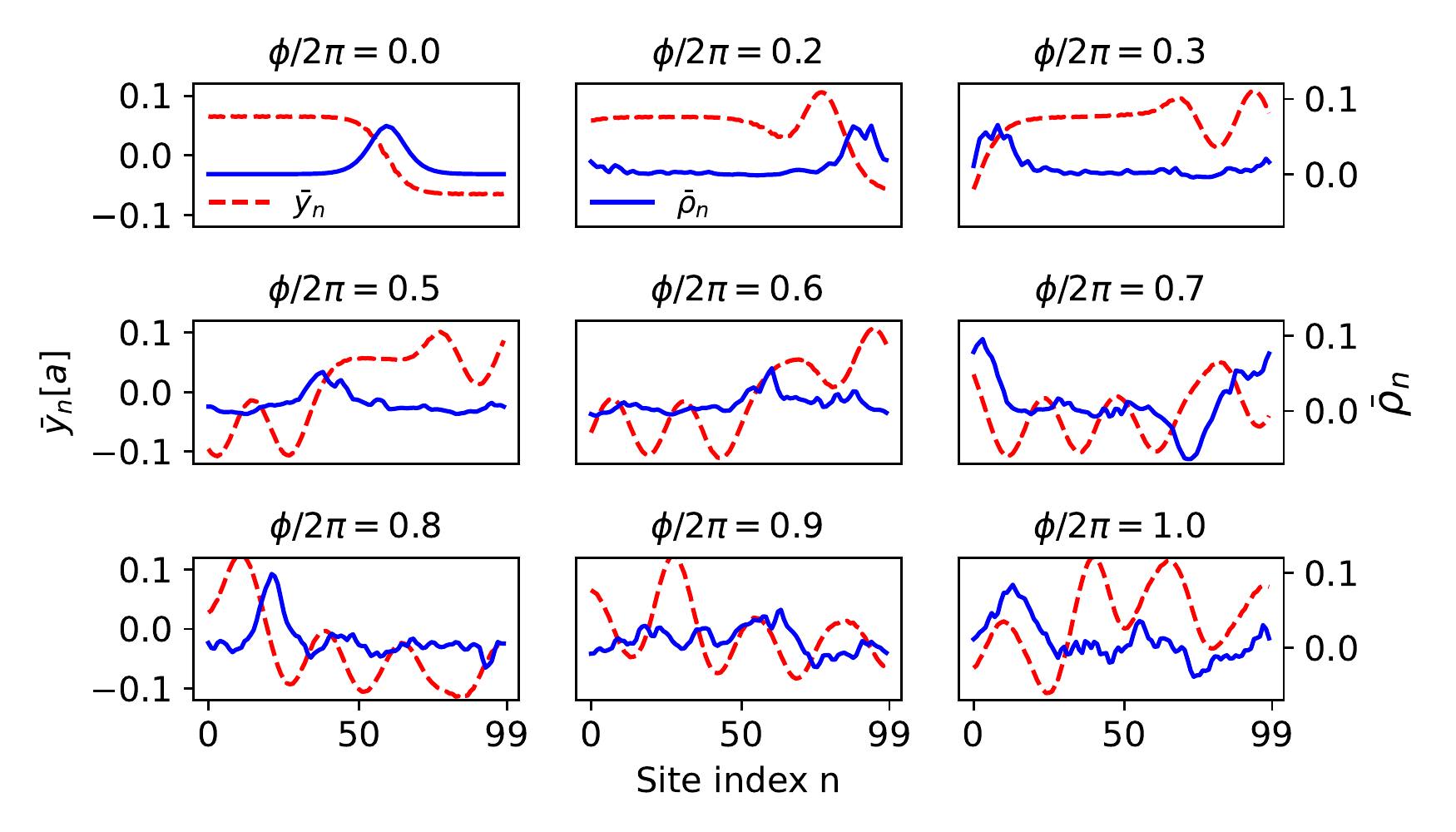}
\caption{ \label{fig:espacio_real_E_p1} 
Staggered lattice deformation field $\bar{y}_n$ (dashed red line) and electronic  excess charge $\bar{\rho}_n$ (continuous blue line)
versus site index $n$ of the tPA chain, computed for an electric field $E=0.1E_0$ (an order of magnitude larger than that of Fig.  \ref{fig:espacio_real_E_p01}). Initially, the soliton moves like a well-defined single object, but for $\phi \gtrsim \pi$ it becomes unstable and breaks down.}
\end{figure*}

Phenomenologically, whenever the solitonic excitation evolves in time as a single stable quantum object the center of mass of the domain wall and the maximum of the electronic excess charge coincide. This can be clearly seen in Fig. \ref{fig:espacio_real_E_p01}, where we simultaneously show  both $\bar{y}_n\left(t\right)$ and $\bar{\rho}_n\left(t\right)$ for different times. These ``snapshots'' correspond to a simulation obtained for a constant and uniform external field $E=0.01E_0$, so that $t_{f}=628 \omega_0^{-1}$. 
At $t = 0$, the deformation field $\bar{y}_n$ shows a kink which connects the two different minima corresponding to the two possible Peierls distortions at equilibrium (i.e., the two possible vacuum states of the continuum field theory). Note that since the system has periodic boundary conditions, $y_n(t)=y_{n+N}(t)$, the staggered lattice field $\bar{y}_n(t)$ must satisfy anti-periodic boundary conditions when $N$ is odd, as in the case of Fig. \ref{fig:espacio_real_E_p01}), and periodic boundary conditions for even $N$. Along the whole simulation, we see that the center of mass of $\bar{\rho}_n(t)$ coincides with the center of the domain wall in $\bar{y}_n(t)$; meaning that the moving soliton excitation evolves as a stable self-organized quasi-particle.  As the system evolves the lattice deformation field develops oscillations due to the excitation of phonon modes with well-defined wavenumber. The dynamical emission of phonons in non-equilibrium situations has been reported before\cite{Su80_Dynamics_of_solitons_in_PA, Bishop84_Breathers_in_Polyacetylene, Guinea84_Dynamics_of_PA_chains, Phillpot87_Dynamics_in_polyyne, ono1990motion}. Interestingly, the presence of these phonons modes does not seem to affect (necessarily) the stability of the soliton in the sense that the excess charge remains rigidly attached to the domain-wall. 

While the  stability of a moving soliton is self-evident in Fig. \ref{fig:espacio_real_E_p01}, we note that the ``matching'' of the center of mass of $\bar{\rho}_n\left(t\right)$ and the center of the domain wall $\bar{y}_n\left(t\right)$ does not represent a rigorous nor a quantitative criterion for soliton stability. To illustrate this point, in Fig. \ref{fig:espacio_real_E_p1} we show a similar calculation for a larger value of the electric field $E=0.1 E_0$. Although in the initial steps the soliton evolves as a well-defined object, it clearly breaks down within a complete evolution of the Peierls phase $\phi=2 \pi$. Several questions naturally arise here: why the soliton appears to be stable for $E=0.01 E_0$, while it breaks down for $E=0.1 E_0$? And precisely \textit{when} the instability occurs?  One of the goals in this work is to obtain a microscopic understanding of the destabilizing mechanisms of moving solitons, and to ``measure'' its stability by the means of a suitable theoretical quantity. This would enable to determine the parameter regime within which a moving soliton behave as a stable self-organized object in a generic out-of-equilibrium situation, therefore complementing the information given by the analytical estimate in Eq. (\ref{eq:stability_condition2}). 

\begin{figure*}[t]
\includegraphics[width=1.1\textwidth]{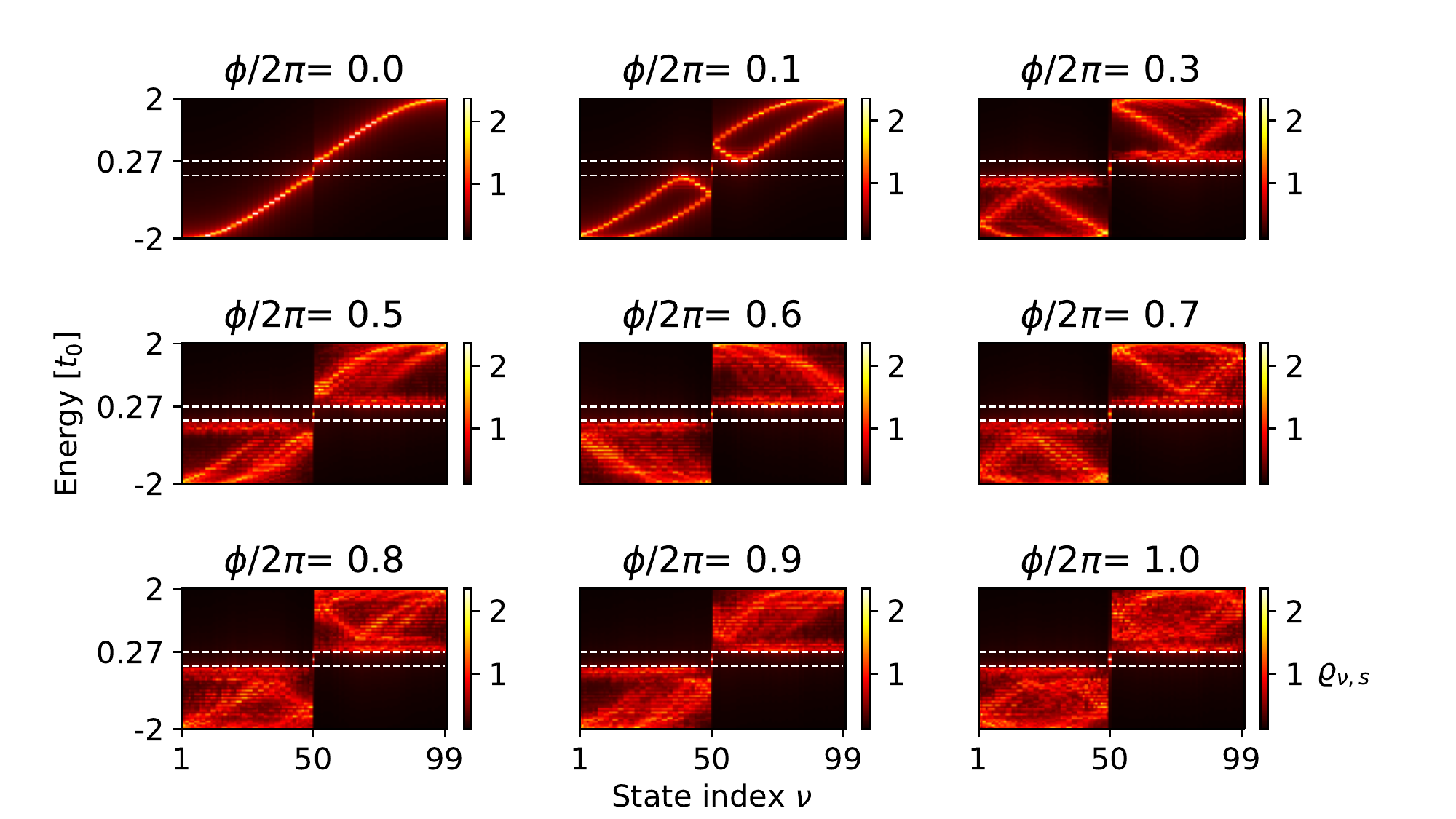}
\caption{Colormaps of the instantaneous Density of States (iDOS) [Eq. (\ref{eq:iDOS})] at different values of the Peierls phase $\phi$. The electric field $E=0.01E_0$ and the values of the Peierls phase are the same as those in Fig. \ref{fig:espacio_real_E_p01}. Each figure in the panel corresponds to a snapshot of the instantaneous electronic structure as a function of the frequency $\omega$ (ordinates) and the instantaneous eigenstate index $\nu$ (abscissas).  The (instantaneous) Peierls gap, delimited by white dashed lines, separates the conduction from the valence bands. The bright spot in the center of the Peierls gap corresponds to the instantaneous zero-energy eigenstate. During the whole simulation both the Peierls gap and the weight of the zero-energy state remain unperturbed, despite the fact that the evolution of bulk states (valence and conduction states) is non-adiabatic.\label{fig:espacio_energia_E_p01}}
\end{figure*}

\begin{figure*}[t]
\includegraphics[width=1.1 \textwidth]{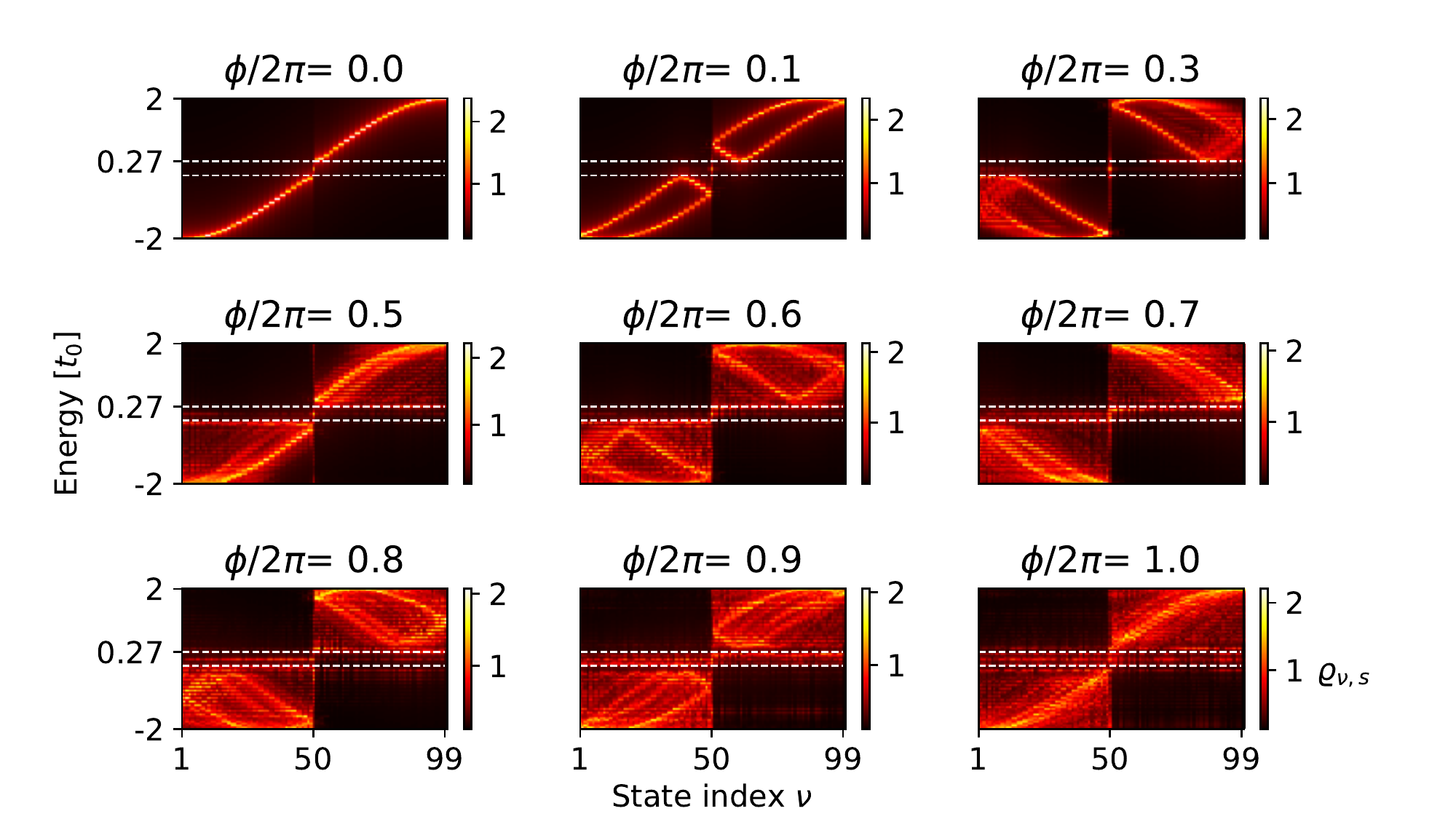}
\caption{\label{fig:espacio_energia_E_p1}
Colormaps of the instantaneous Density of States (iDOS) [Eq. (\ref{eq:iDOS})] computed for $E=0.1E_0$, and for the same values of the Peierls phase as in Fig. \ref{fig:espacio_real_E_p1}. In contrast to the situation in Fig. \ref{fig:espacio_energia_E_p01} (computed for $E=0.01 E_0$), here after some time the Peierls gap becomes populated with new states originated in large amplitude oscillations in staggered field $\bar{y}_n$. As a result, the gap becomes poissoned and the adiabatic condition Eq. (\ref{eq:adiabatic_criterion}) for the zero-energy state  is eventually violated. The evolution of the zero-energy state becomes non-adiabatic and the soliton becomes unstable.}
\end{figure*}

To fix ideas, it is instructive to study the evolution of the single-particle dynamical states of the problem. To that end we define an ``instantaneous'' spectral density of states (iDOS)
\begin{align}\label{eq:iDOS}
\varrho_{\nu,s}(t,\omega) &=-\frac{1}{\pi}\text{Im}\ \langle\psi_{\nu,s}(t)|\hat{G}^{r}(\omega, t)|\psi_{\nu,s}(t)\rangle \nonumber \\
&=-\frac{1}{\pi}\sum_{\nu^\prime,s}\frac{|\langle\psi_{\nu,s}(t)|\phi_{\nu^\prime,s}(t)\rangle|^2\delta}{\left(\hbar\omega-\epsilon_{\nu^\prime,s}(t)\right)^2+\delta^2}
\end{align}
where the dynamical state $|\psi_{\nu,s}(t)\rangle$ is calculated using Eq.  (\ref{eq:wavefunction_evolution}), and where the operator $\hat{G}^{r}(\omega, t)$ is the instantaneous retarded Green's function, defined in the complex frequency plane $z$ as
\begin{align}\label{eq:inst_green}
\hat{G}^{r}(z, t) &\equiv \sum_{\nu^\prime,s^\prime} \frac{|\phi_{\nu^\prime,s^\prime}(t)\rangle \langle \phi_{\nu^\prime,s^\prime}(t)|}{z-\epsilon_{\nu^\prime,s^\prime}(t)},
\end{align}
written in terms of the instantaneous eigenvalues and eigenvectors of $H_\text{SSH}(t)$, i.e.,  $\epsilon_{\nu^\prime,s}(t)$ and $|\phi_{\nu^\prime,s}(t)\rangle$, respectively. A small imaginary part $\delta$ has been introduced in Eq. (\ref{eq:iDOS}) in order to avoid the non-analyticities given by the instantaneous spectrum in the real $\omega$-axis.

In addition to giving information about the occupancy $n_{\nu^\prime,s}(t)$ of the instantaneous eigenstate $|\phi_{\nu^\prime,s}(t)\rangle$, intuitively speaking the iDOS provides a ``measure'' of  adiabaticity in the system: if a given dynamical state $|\psi_{\nu,s}(t)\rangle$ evolves adiabatically from the eigenstate $|\psi_{\nu,s}(0)\rangle$, then there will be a one-to-one correspondence given by  Eq. (\ref{eq:adiabatic_theorem}) with the instantaneous eigenstate $|\phi_{\nu^\prime,s}(t)\rangle$  \textit{at all times}. Quantitatively, this correspondence will be measured by the projection $|\langle\psi_{\nu,s}(t)|\phi_{\nu^\prime,s}(t)\rangle|^2$ in the numerator of Eq. (\ref{eq:iDOS}), and deviations of this quantity from the Kronecker delta $\delta_{\nu,\nu^\prime}$ can therefore be interpreted 
as a breaking of adiabaticity.

In Fig. \ref{fig:espacio_energia_E_p01}, we show the iDOS $\varrho_{\nu,s}(t,\omega)$ at different times, and for the same parameters as in Fig. \ref{fig:espacio_real_E_p01}. At each time (which coincide with those in Fig. \ref{fig:espacio_real_E_p01}), we show colormaps of $\varrho_{\nu,s}(t,\omega)$ as a function of frequency $\omega$ (vertical axis) for each of the evolved states (label $\nu$ in the horizontal axis). Therefore, each plot is a snapshot of the full evolved spectrum. At $t=0$, the state $|\psi_{\nu,s}(t)\rangle$ and the instantaneous eigenstate $|\phi_{\nu^\prime,s}(t)\rangle$ coincide (i.e., $|\langle\psi_{\nu,s}(t)|\phi_{\nu^\prime,s}(t)\rangle|^2=\delta_{\nu,\nu^\prime}$) and, as expected, the one-to-one correspondence is explicit in Fig.  \ref{fig:espacio_energia_E_p01}. Note in this figure the presence of the Peierls gap from $\omega=-0.27 t_0$ to $\omega=0.27t_0$, separating valence and conduction bands. In addition, note the presence of the zero-energy state (single bright dot at $\omega=0$) which appears as a consequence of the domain-wall in the lattice distortion field $\bar{y}_n$. Since the instantaneous eigenstates $|\phi_{\nu,s}(t)\rangle$ have been sorted as a function of increasing instantaneous eigenvalues $\epsilon_{\nu,s}(t)$, the first 49 states in the horizontal axis correspond to the valence band, while the last 49 states form the conduction band. The zero-energy state is therefore the instantaneous eigenstate with $\nu=50$.

At times $t>0$, the non-adiabatic time evolution becomes apparent for the bulk states (as mentioned previously), which spectrally decompose into several instantaneous eigenstates. Roughly speaking, in the absence of electron-phonon interaction (i.e., parameter $\alpha=0$ in Eq. \ref{eq:hamiltonian}), the expected time evolution of the valence- and conduction-band $k$-states can be understood in simple terms according to the semiclassical equation of motion $\dot{k}=-eEt/\hbar$ \cite{ashcroft}, and a single valence or conduction band is expected.  However, for finite $\alpha$ the excitation of a phonon mode with momentum $q$ produces ``replicas'' of the spectrum due to the interaction of states with momentum $k$ and $k+q$, and a non-trivial electronic structure with several new anti-crossings emerges. After some time, the original one-to-one correspondence at $t=0$ is completely lost for the bulk states. Note that this breaking of adiabaticity does not imply the generation of electron-hole excitations in the system, as is evident from the lack of population in the diagonal blocks in Fig. \ref{fig:espacio_energia_E_p01}. In other words: valence-band states only mix with among themselves and similarly for valence-band states. This is in good agreement with the low-field conditions given in Eqs. (\ref{eq:stability_condition2}) and (\ref{eq:electric_breakdown}). 0.
 
Note that while adiabaticity is lost for the bulk states, the exception to this picture is precisely the zero-energy state, for which the projection onto the zero-energy instantaneous state remains close to 1 (see Fig. \ref{fig:projection_psi0} for more details).

\begin{figure}[t]
\includegraphics[scale=0.5]{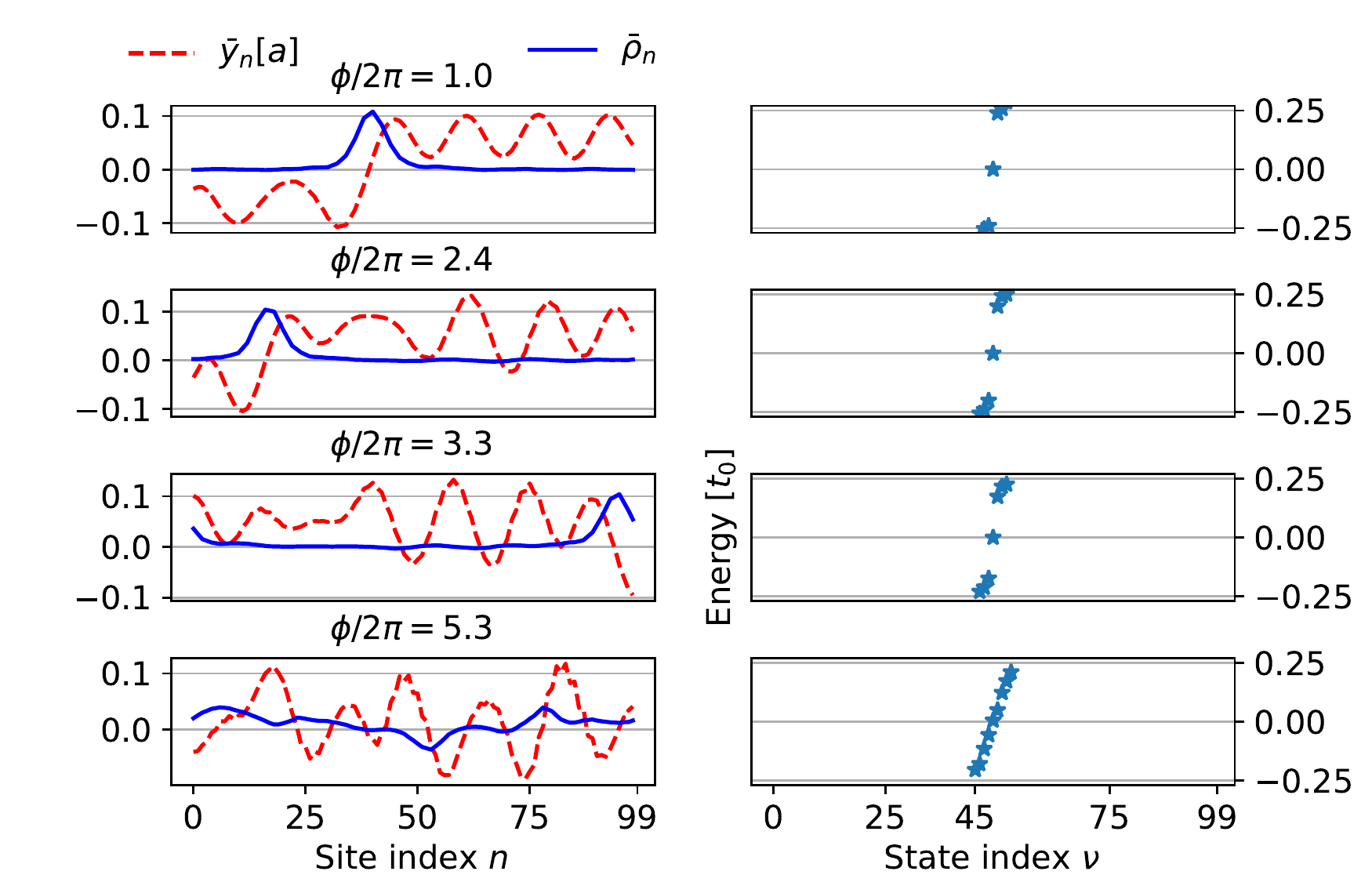}
\caption{(Left panel) Staggered lattice deformation field $\bar{y}_n$ (dashed red line) and electronic excess charge  $\bar{\rho}_n$ (continuous blue line) versus site index $n$, computed for $E=0.03E_0$ and at different values of the Peierls phase. (Right panel) Energy spectrum of instantaneous in-gap states versus eigenstate index $\nu$, computed for the same values of the Peiers phase. As the system evolves, the dynamical instability evidenced in the real-space quantities correlates with the closing of the Peierls gap.}\label{fig:soliton_broken_small_field}
\end{figure}

On the other hand, in Fig. \ref{fig:espacio_energia_E_p1} we present the same calculation for $E=0.1 E_0$ as in  Fig. \ref{fig:espacio_real_E_p1}, for which the moving soliton solution breaks down within a period $2\pi$ of the Peierls phase $\phi$. We therefore ask the question: how does this unstability ``looks'' in the instantaneous spectrum? Interestingly, in addition to the different evolution of the bulk states (i.e., a more ``diabatic'' evolution due to the larger field $E$, and the hint of some electron-hole excitations, there is a striking difference with respect to Fig. \ref{fig:espacio_energia_E_p01} concerning the gap region. Note that after some time ($t\simeq 46.3 \omega_0^{-1}$) the original gap becomes populated with new states, which were absent before. The emergence of these new in-gap states can be rationalized in the following way: as a consequence of the large amplitude oscillations in the staggered field $\bar{y}_n(t)$ (i.e., large oscillations of the phonon mode) of the order of $\sim y_0$, new kink-antikink pairs are produced along the chain, and new low-energy electronic states appear. When these kink-antikink pairs proliferate, they generate more and more low-lying electronic states which ``poisson''  the gap $\Delta_g$, effectively reducing it and destabilizing the topological protection of the zero-energy state. This is more clearly seen in Fig. \ref{fig:soliton_broken_small_field}, where we show both the configuration of the lattice deformation field $\bar{y}_n(t)$ and the electronic excess charge $\bar{\rho}_n(t)$ (left panel) along with the corresponding instantaneous spectrum of in-gap states. Note that this connection between real-space maps and the (instantaneous) energy spectrum in the adiabatic regime provides a lot of information and, to the best of our knowledge, has not been explored before. Interestingly, the results in Fig. \ref{fig:soliton_broken_small_field} have been obtained for a field $E=0.03E_0< E_\text{c}$, which is sufficiently small for the instability to appear at very long times (i.e., the simulation time is $\phi=[0,12\pi]$), but which is nevertheless large enough for the instability to eventually occur.

Building upon these ideas, we now propose that the quantity
\begin{align}\label{eq:p0}
p_0(t)&=|\langle\psi_{0}(t)|\phi_{0}(t)\rangle|^2,
\end{align}
i.e., the projection of the evolved zero-energy state (here for simplicity the state $\nu=50$ is indicated by the subindex ``0'') onto the instantaneous zero-energy state, is a \textit{bona fide} measure of the stability of a dynamical soliton. In Fig. \ref{fig:projection_psi0} we show $p_0(t)$  as a function of time (measured in terms of the Peierls phase $\phi$) for several values of the external field $E$. We note that $p_0(t)$ remain stable and close to 1 for $E=0.03\ E_0$ for the most of the simulation, but eventually drops and deviates from 1 around $\phi \approx 10 \pi$. The onset of this deviation (indicated as black spots A, B, C for each value of $E$) is relatively abrupt and coincides with the onset of the instability observed in the real space maps of $y_n(t)$ and $\bar{\rho}_n(t)$ (see Fig. \ref{fig:soliton_broken_small_field}). Therefore, we postulate that $p_0(t)$ can be used as a simple indicator of the stability of dynamical solitons in the SSH model. To substantiate this claim, we present in Fig. \ref{fig:instability} the real space maps for both $\bar{y}_n(t)$ and $\bar{\rho}_n(t)$ \emph{before} and \emph{after} the instability of $p_0(t)$ (black dots) for each value of the external field. Interestingly, the excess charge before the instability in $p_0(t)$ is more or less concentrated around the original domain-wall and retains the main features of a stable moving soliton. However, after $p_0(t)$ falls below  $p_0(t)\lesssim 0.8$ the excess charge spreads over the whole system and the soliton breaksdown.

\begin{figure}[t]
\includegraphics[scale=0.55]{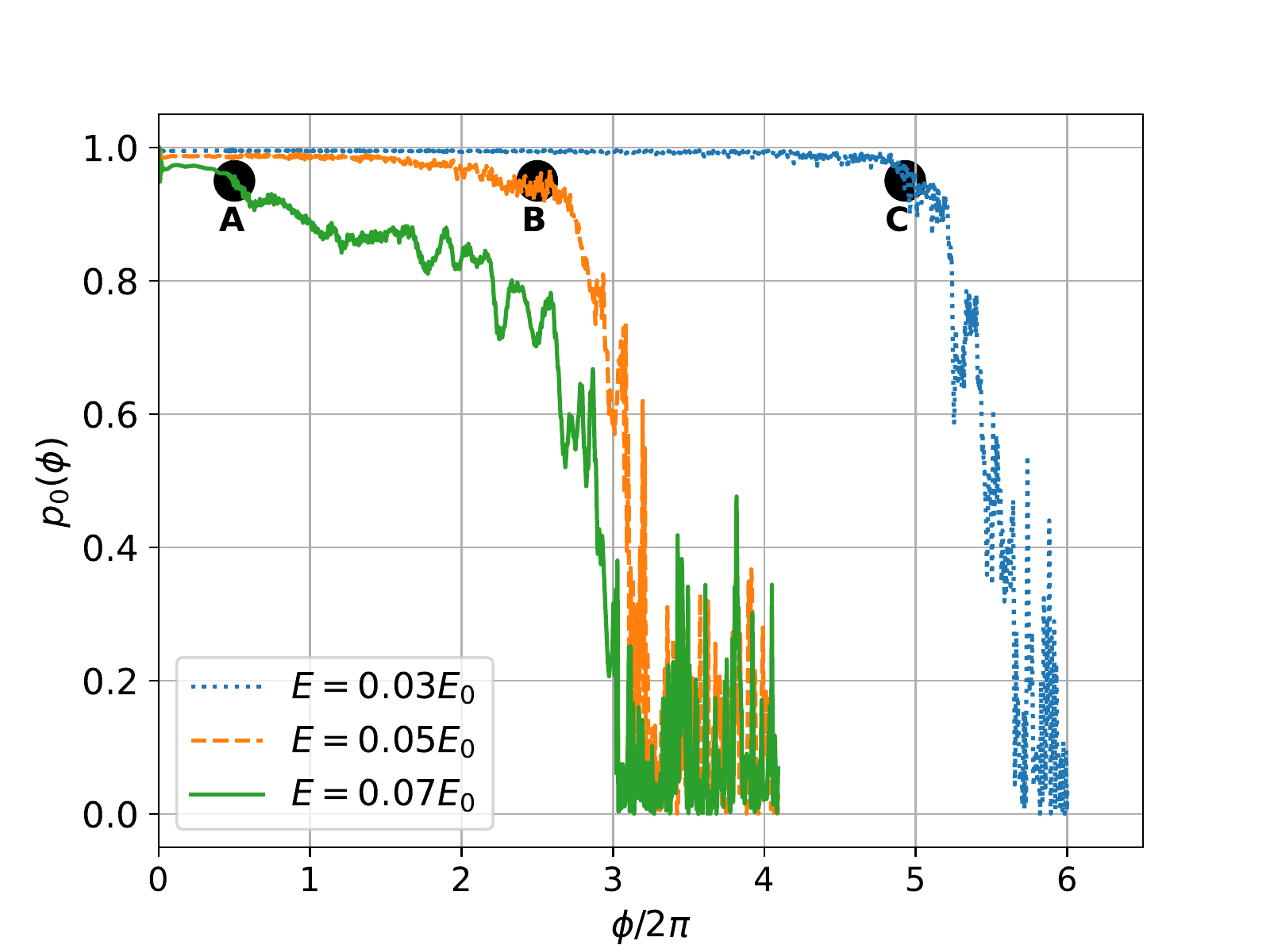}
\caption{Projection $p_0(t)$ [Eq.(\ref{eq:p0})] versus the Peierls phase $\phi$ for different values of $E$. A stable evolution of the soliton excitation as a single quantum object is evidenced in $p_0(t)\simeq 1$. After some time (which strongly depends on the value of $E$), the value of $p_0(t)$ suddenly drops, meaning that the evolution of the zero-energy state becomes non-adiabatic (see black dots indicated A, B and C). This drop in $p_0(t)$ is correlated with the soliton breakdown manifested in the real-space variables  $\bar{y}_n$ and $\bar{\rho}_n$ (see Figs. \ref{fig:soliton_broken_small_field}) and Fig. \ref{fig:instability}, and allows to consider $p_0(t)$ as a reliable quantitative indicator of the stability of out-of-equilibrium solitonic excitations.\label{fig:projection_psi0}}
\end{figure}

\begin{figure}
\includegraphics[scale=0.5]{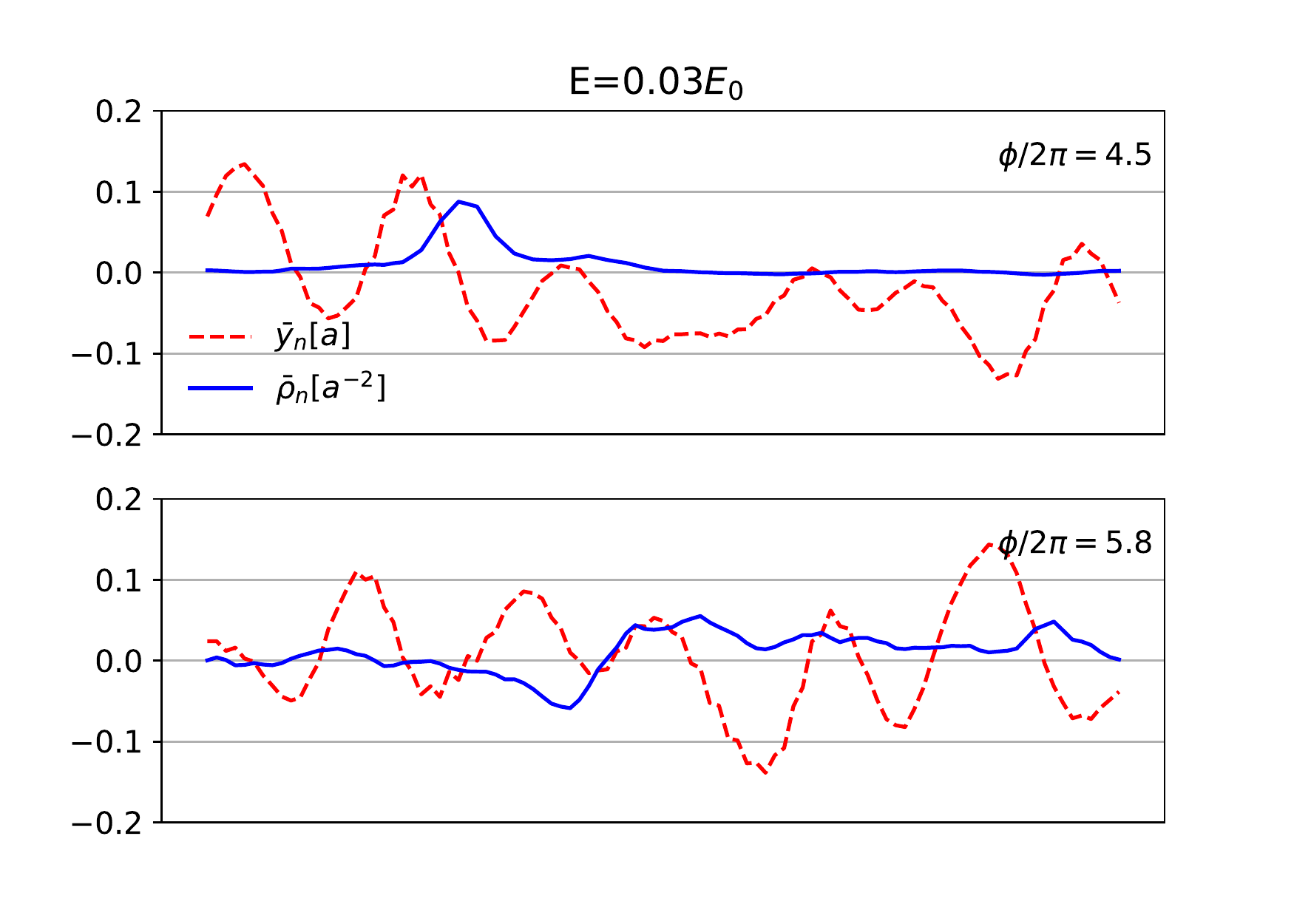}
\includegraphics[scale=0.5]{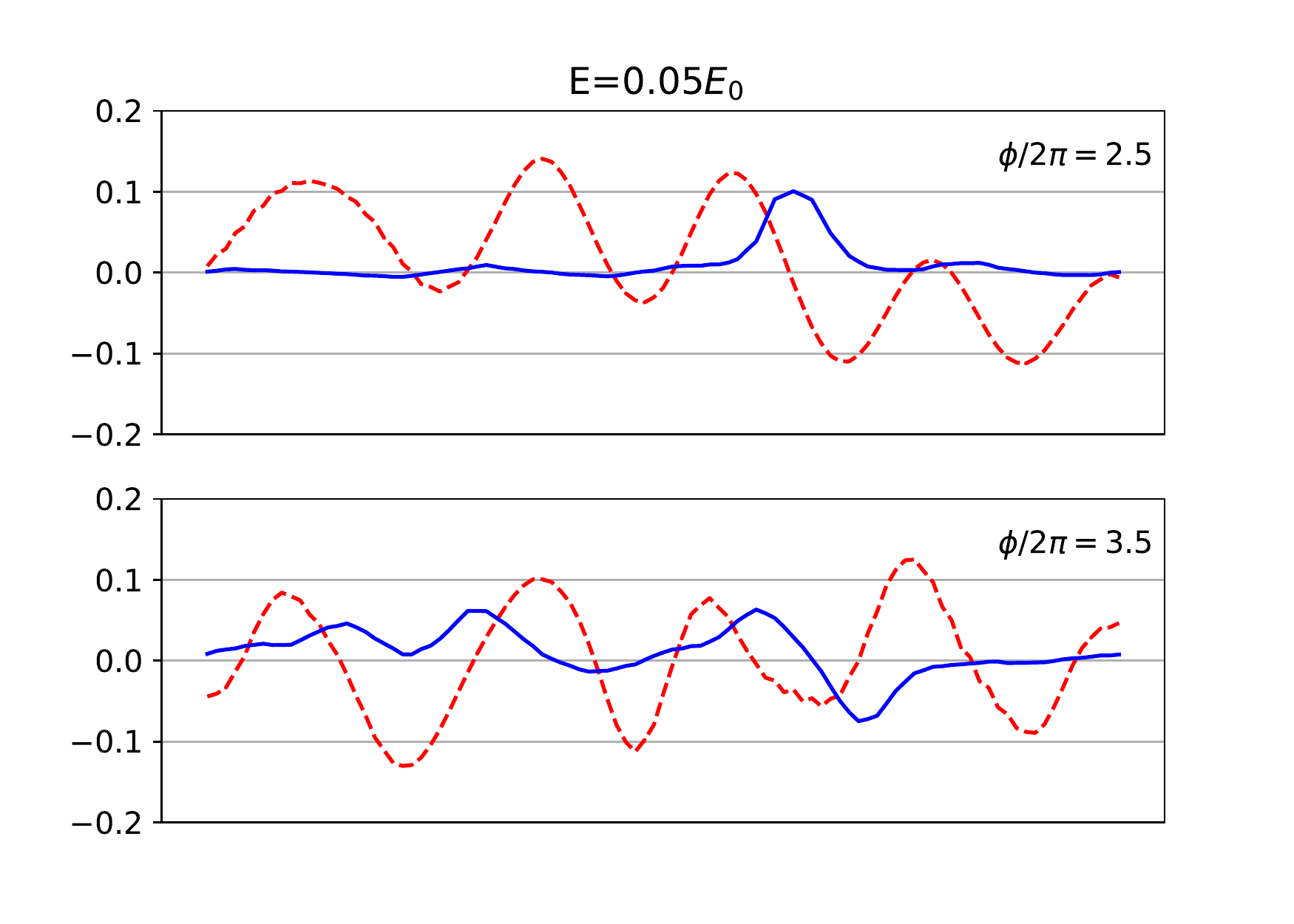}\\
\includegraphics[scale=0.55]{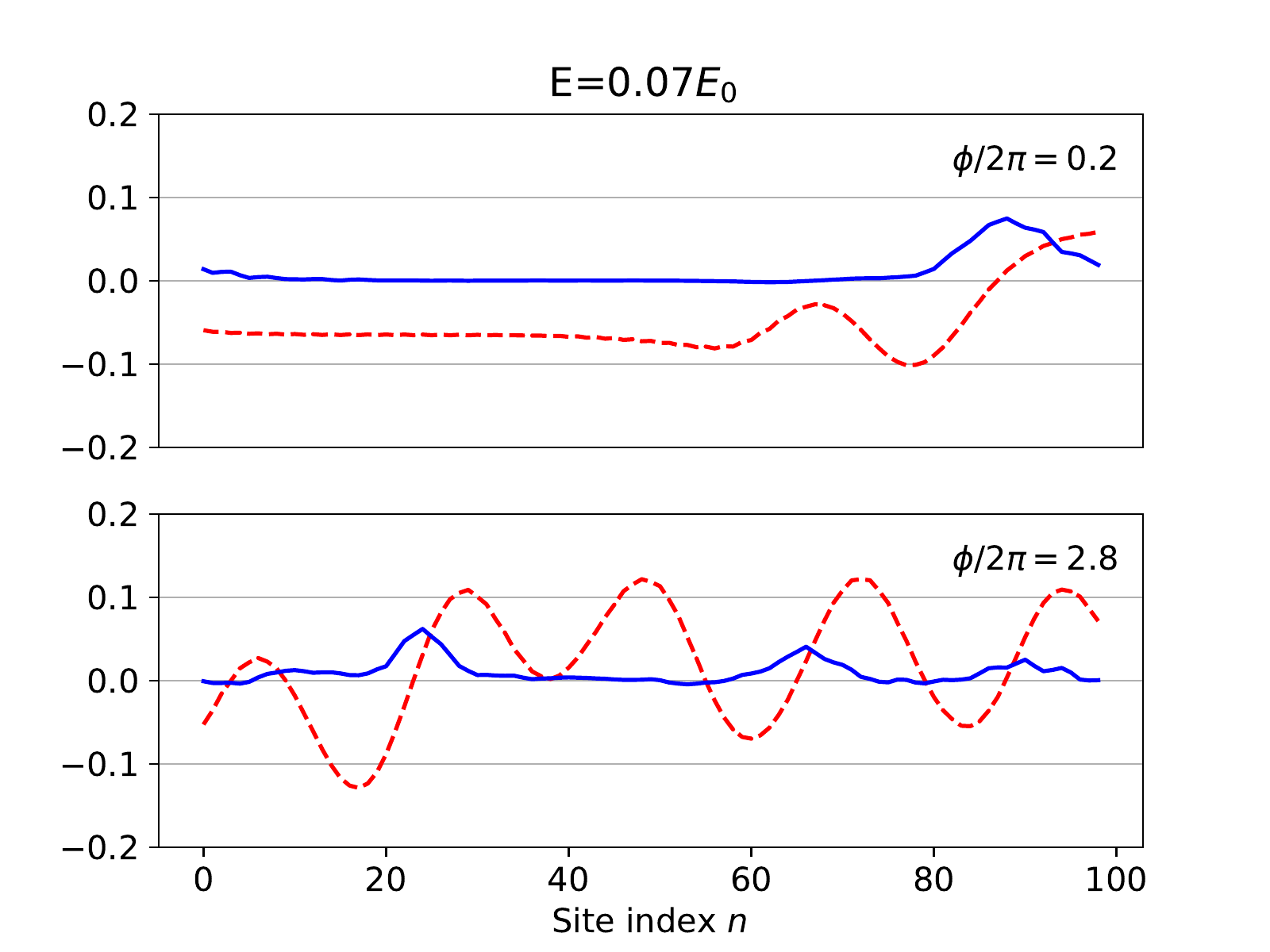}
\caption{Staggered lattice deformation field $\bar{y}_n$ (dashed red lines) and electronic excess charge $\bar{\rho}_n$ (continuous blue lines) versus site index $n$, computed for different values of constant electric field $E$. For each value of $E$ we evaluate the fields $\bar{y}_n$ and $\bar{\rho}_n$ before and after the instabilities marked with black dots in Fig. \ref{fig:projection_psi0} (see the corresponding insets for details). An evident change of behavior (i.e., delocalization of the electronic excess density and larger amplitude in the lattice deformation field) occurs in the figures after the instability.}\label{fig:instability}
\end{figure}



\subsection{Saturation velocity analysis}

In this section we focus on the analysis of the saturation velocity of the moving soliton. 
When a charged soliton is accelerated in an external field, the interaction with acoustic phonons is known to introduce friction \cite{ono1990motion, ono1991motion, Kuwabara92_Damping_of_soliton_velocity}. This damping takes place even at zero temperature and in the absence of thermally excited phonons, since the phonon emission is originated in the motion of the soliton itself. Therefore, after a characteristic time $\tau$ the soliton eventually attains a saturation velocity $v_\text{sat}$\cite{ono1990motion}. Alternatively, if the field is suddenly turned off, its velocity relaxes to $v_\text{sat}\rightarrow 0$. According to Ref. \onlinecite{Kuwabara92_Damping_of_soliton_velocity}, this relaxation time is minimal (i.e., friction is therefore more efficient) when the initial velocity $v$ of the soliton is approximately equal to the sound velocity of acoustic phonons in tPA, $v_s$. In particular, in Ref. \onlinecite{ono1990motion}, Ono and Terai  have shown this friction effect by computing the position of the center of mass of the excess charge distribution $\bar{\rho}_n(t)$ as a function of $t$, i.e., 
\begin{align}\label{eq:center_of_mass}
x_c(t)&=\frac{Na}{2\pi}\tan^{-1}\left(\frac{\sum_n \bar{\rho}_n(t) \sin\theta_n}{\sum_n \bar{\rho}_n(t) \cos\theta_n}\right),
\end{align}
with $\theta_n=2\pi n/N$, and from here the velocity of the soliton is obtained taking the numerical derivative. Note that the definition (\ref{eq:center_of_mass}) is consistent with the periodic boundary conditions imposed in the system. 

\begin{figure}
\includegraphics[scale=0.5]{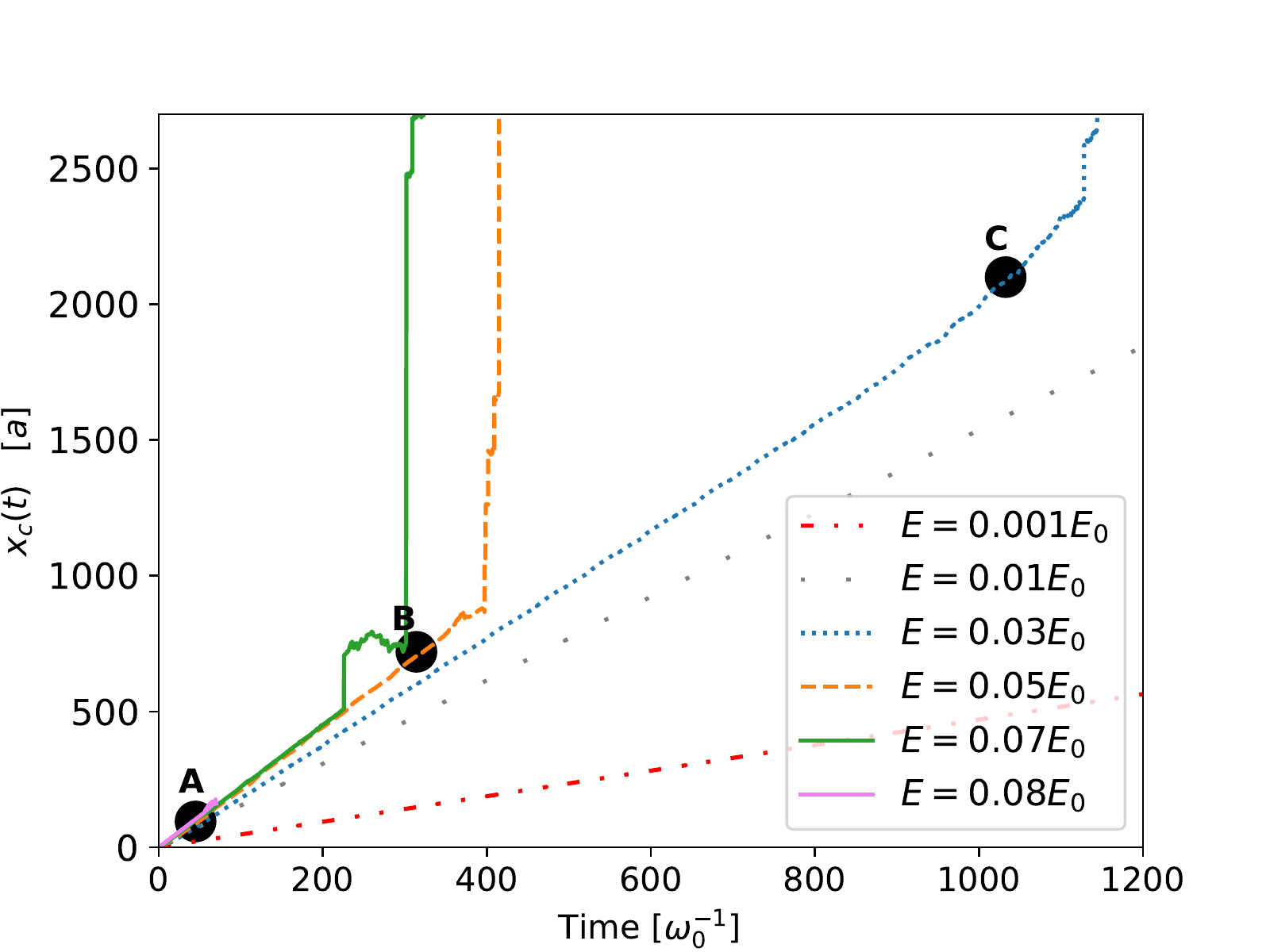}
\caption{Center of mass of the electronic excess charge, $x_c(t)$ [Eq. (\ref{eq:center_of_mass})], as a function of time. After a while, the behavior of $x_c(t)$ shows a linear behavior in $t$, indicating the presence of damping due to the interaction with phonons (see Ref. \onlinecite{Kuwabara92_Damping_of_soliton_velocity}). The breakdown of the soliton is manifested as an abrupt upturn of the curves. For comparison, we also show the occurrence of the instabilities in Fig. \ref{fig:projection_psi0}, obtained from the analysis of $p_0(t)$. In all cases, the behavior of $p_0(t)$ is able to predict the breakdown of the soliton, and in some cases (e.g., for $E=0.07 E_0$) the parameter $p_0(t)$ is actually a more reliable indicator of the soliton instability.\label{fig:centro_masa_vs_E}}
\end{figure}

One conclusion in Ref. \onlinecite{ono1990motion} is that the velocity of the soliton reaches a saturation value $v_\text{sat}$ which is \emph{independent} of the value of the applied external field $E$. We believe that this conclusion is unphysical because it cannot reproduce the necessary requirement that $v_\text{sat} \rightarrow 0$ in the limit $E\rightarrow 0$. On the other hand, at sufficiently large fields the soliton eventually breaksdown and it is no longer a meaningful concept. 

Therefore, here we revisit the issue of the saturation velocity in the context of our previous stability analysis. We computed the center-of-mass position as a function of time as in  Ref. \onlinecite{ono1990motion}, for different values of the external field. Our results are shown in Fig. \ref{fig:centro_masa_vs_E}. All the curves show, after a time which depends on the particular value of $E$, a clear linear dependence which is indicative of the saturation of the soliton velocity, as expected from the previous considerations. In addition, note the abrupt upturn deviation from linearity in the plots corresponding to $E\geqslant 0.03\ E_0$, which can be associated (but as we show below, does not correspond) to the soliton breakdown. In order to compare with our results, here we have indicated the occurrence of the same instabilities observed in Fig. (\ref{fig:projection_psi0}) in $p_0(t)$  (black spots A, B, and C) for each curve. Quite surprisingly, we note that the instability in $p_0(t)$ always occurs \emph{before} the upturn in each of the curves $x_c(t)$. Moreover,  the value of $x_c(t)$ computed from Eq. (\ref{eq:center_of_mass}) remains well defined and continuous \textit{even after the instability in $p_0(t)$}. In the case of $E=0.07\ E_0$, this behavior even continues for a quite significant part of the simulation, and therefore it becomes meaningless to refer to the ``soliton velocity'' in this regime. A possible explanation for this phenomenon is that the position of the center of mass $x_c(t)$, being an averaged quantity, is less sensitive to the variations in the excess charge $\bar{\rho}_n(t)$ and therefore cannot capture the true soliton breakdown. One conclusion of this analysis is that the behaviour of $x_c(t)$ vs $t$ cannot be taken as an indicator of the soliton stability, and therefore must be complemented with additional information (e.g., with that of $p_0(t)$). This point has been overlooked in previous works. 

\begin{figure}[t]
\begin{flushleft}
\includegraphics[scale=0.6]{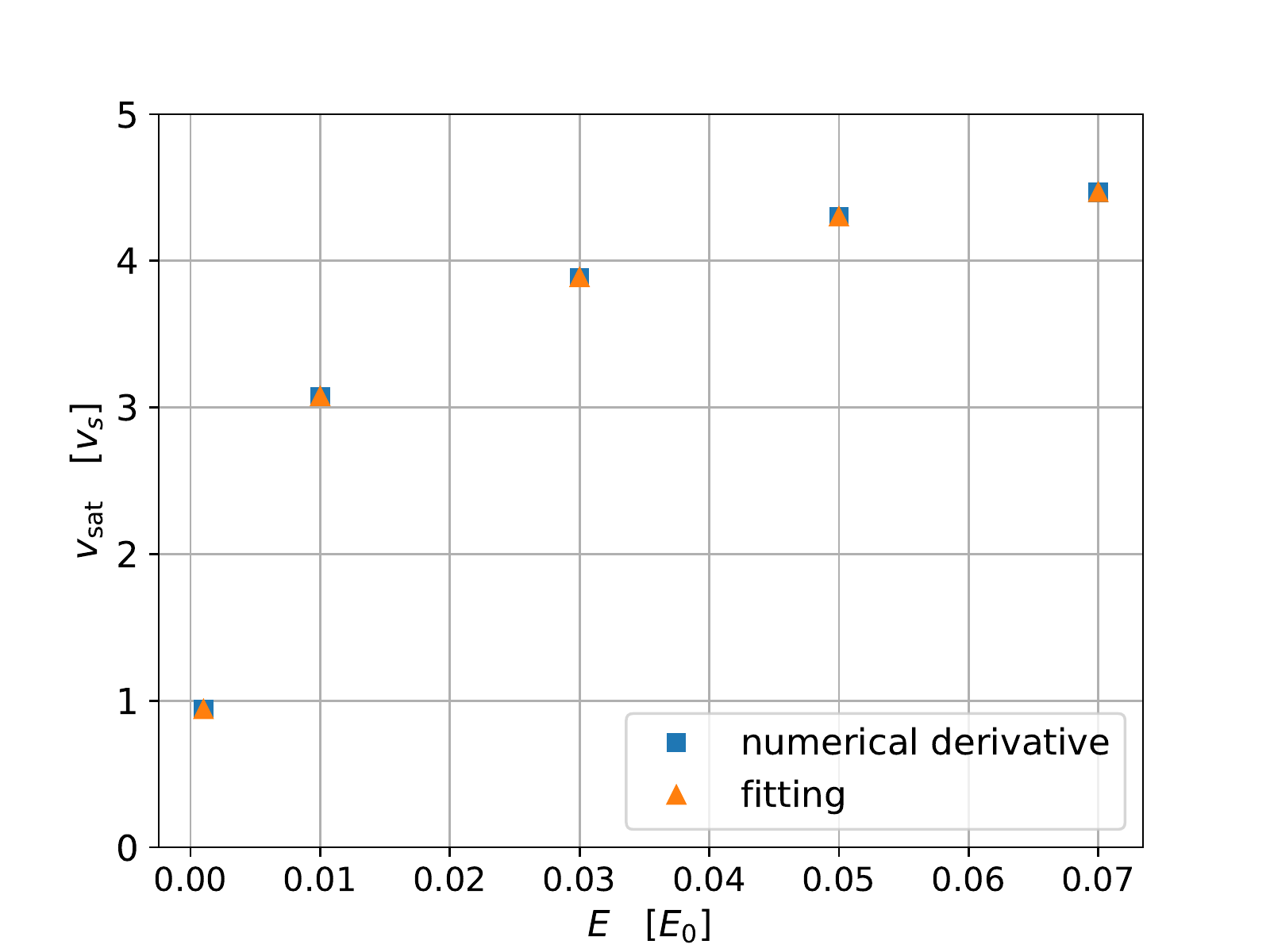}
\end{flushleft}
\caption{Saturation velocity of the soliton, $v_\text{sat}$, computed from the results in Fig. \ref{fig:centro_masa_vs_E} using different numerical methods (see text). The behavior of $v_\text{sat}$ rapidly saturates to a value $\sim 4v_s$, with $v_s$ the sound velocity of acoustic phonons, in agreement with Ref. \onlinecite{ono1990motion}. However, at variance with that work, we reproduce the physically required behavior $v_\text{sat} \to 0$ in the limit $E\to 0$. \label{fig:velocidad_saturacion_vs_E}}
\end{figure}

From the data points in Fig. \ref{fig:centro_masa_vs_E}, we have obtained the saturation velocity via two methods: a) by taking the average of the numerical derivative when it reaches the saturation regime (this method is similar to Ref. \onlinecite{ono1990motion}), and b) by fitting the data of Fig. \ref{fig:centro_masa_vs_E} with the formula: 
\begin{equation}
x_c(t) = p_1 \left(p_2 + t^2 \right)^{p_3} + p_4
\end{equation}
where $p_1,p_2,p_3,p_4$ are fitting parameters, obtained by least squares method. Here $p_3$ is approximately $1/2$ in order for $x_c(t)$ to reproduce a linear behavior at large $t$. Once we obtain these parameters, we take the analytical derivative and take the limit $t\to\infty$. This is numerically more robust and less noisy than the numerical derivative implemented in 
Ref. \onlinecite{ono1990motion}. 
Our results are shown in Fig. \ref{fig:velocidad_saturacion_vs_E}. Note that, at variance with that work, we obtain a  saturation velocity which actually \emph{depends on the applied field}. Our numerical results, specially those corresponding to very low fields, have been checked against very long simulations corresponding to a total Peierls phase $\phi(t_\text{max})=6 \pi$ (i.e., 40-50 times longer than  the time required to reach $v_\text{sat}$, given by the empiric formula (4.5) in Ref. \onlinecite{ono1990motion})
\begin{align}\label{eq:Ono_Terai_formula}
\tau\omega_{0}&=-17.1\ln\left(\frac{E}{E_{0}}\right)-45.1.
\end{align}
In this way, we are sure we are well in the saturated regime. In addition, the convergence of these curves has been checked for different values of $\Delta t =0.01\omega_0^{-1}, 0.001\omega_0^{-1}, 0.0001\omega_0^{-1}$. For $\Delta t =0.01\omega_0^{-1}$ all of the computed quantities converge within a tolerance of $0.2\%$. Our results are indeed consistent with the physically intuitive limit $v_\text{sat} (E\rightarrow 0) \rightarrow 0$. However, in the regime $E>0.03\ E_0$, we note a rapid convergence to  a constant value which is approximately $v_\text{sat} \sim 4 v_s$, as obtained by Ono and Terai. It is probably this rapid saturation which explains the conclusions obtained by those authors.
  
\section{Analysis of finite-size effects}\label{sec:finite_size}

In this last section we study the effects on the stability of the moving soliton due to the finite size $N$ of the system. Generically speaking, since the soliton is a spatially-localized excitation within a length $\xi$,  equilibrium properties (i.e., in the absence of external field $E=0$) are not expected to change as long as the system size $L=Na$ is much larger than the width of the soliton $\xi$. However, the case of moving solitons is quite different, as we show below. 

\begin{figure}[h!]
\includegraphics[scale=0.5]{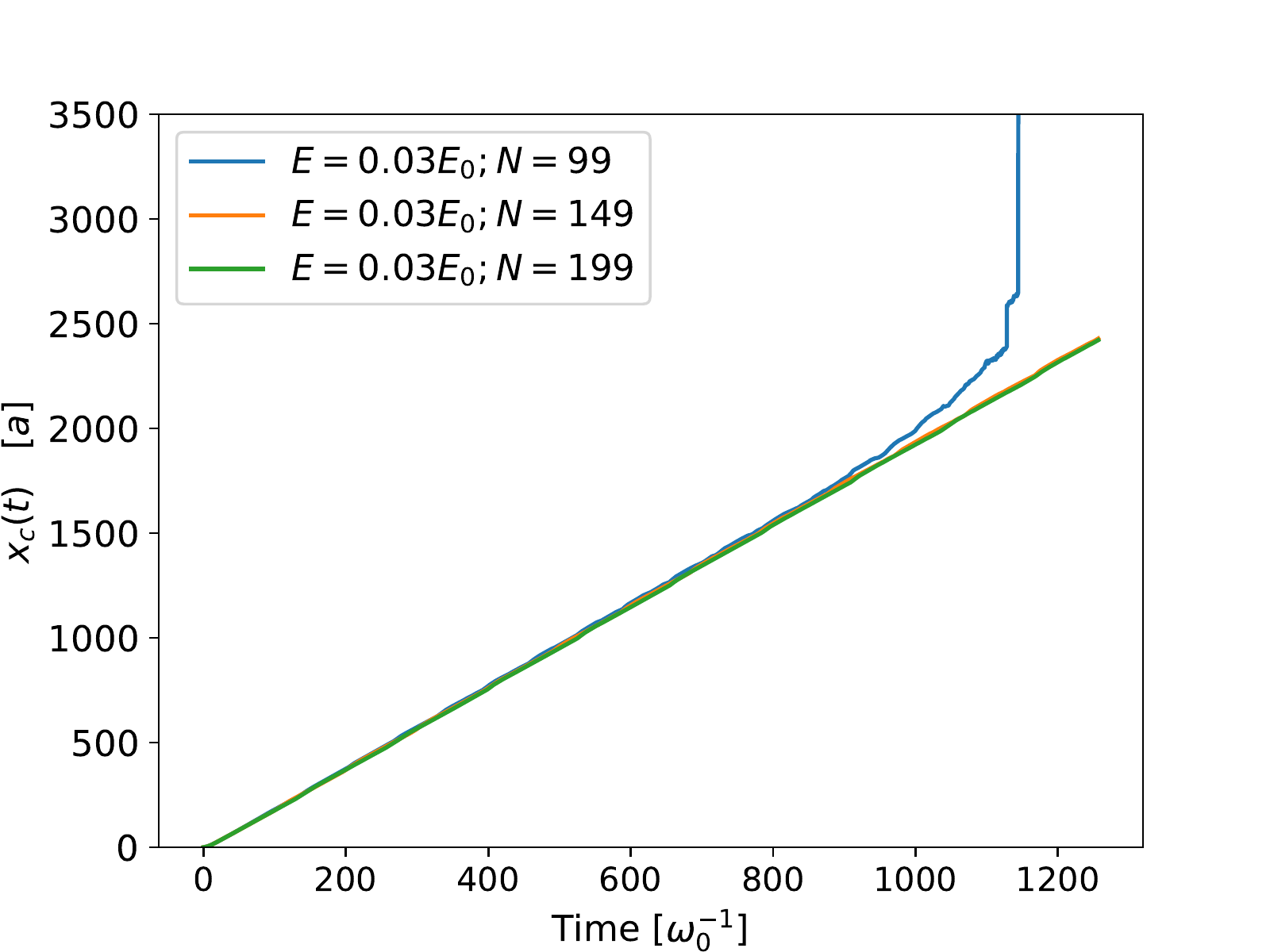}\\
\includegraphics[scale=0.5]{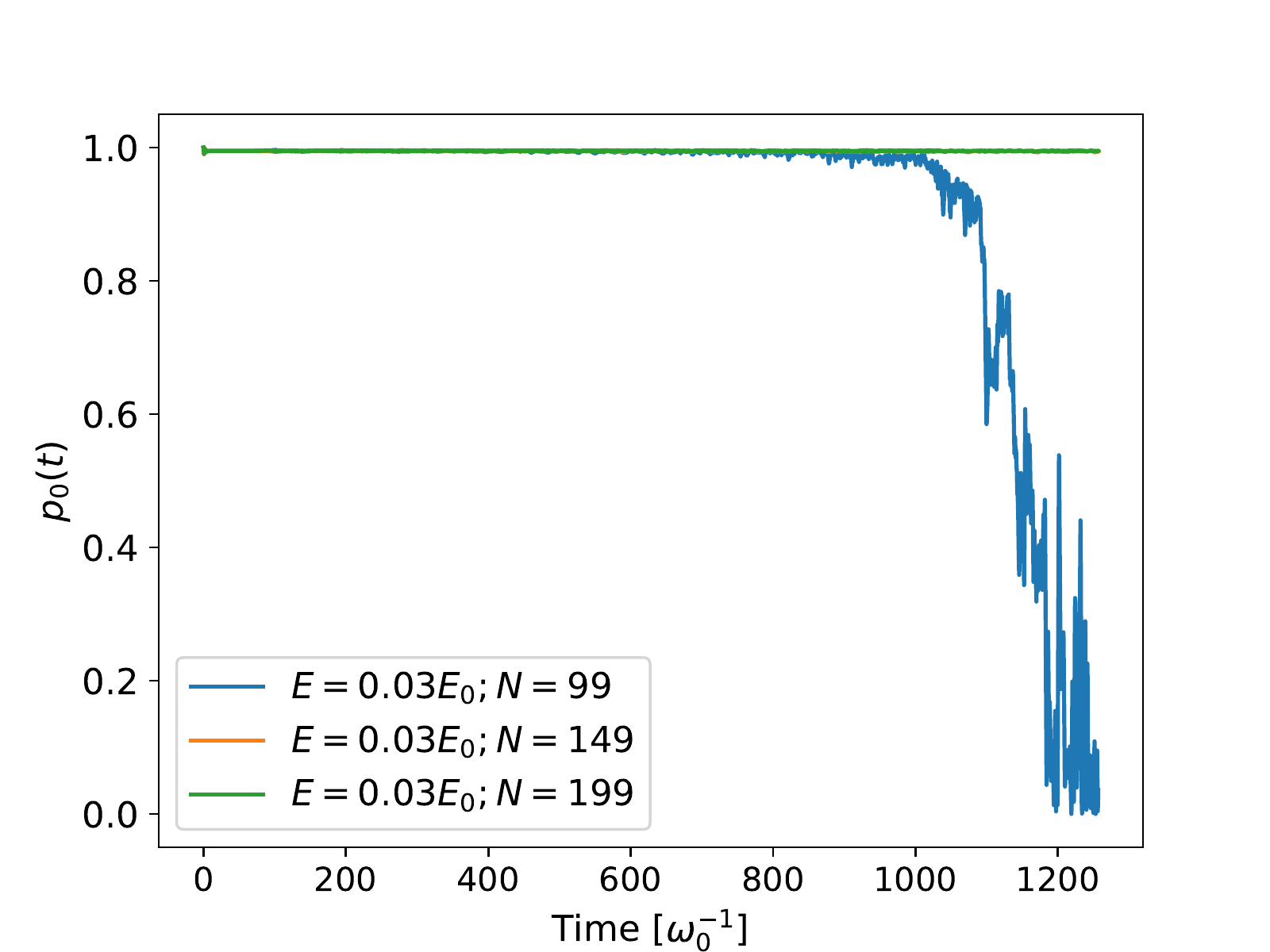}\\
\includegraphics[scale=0.5]{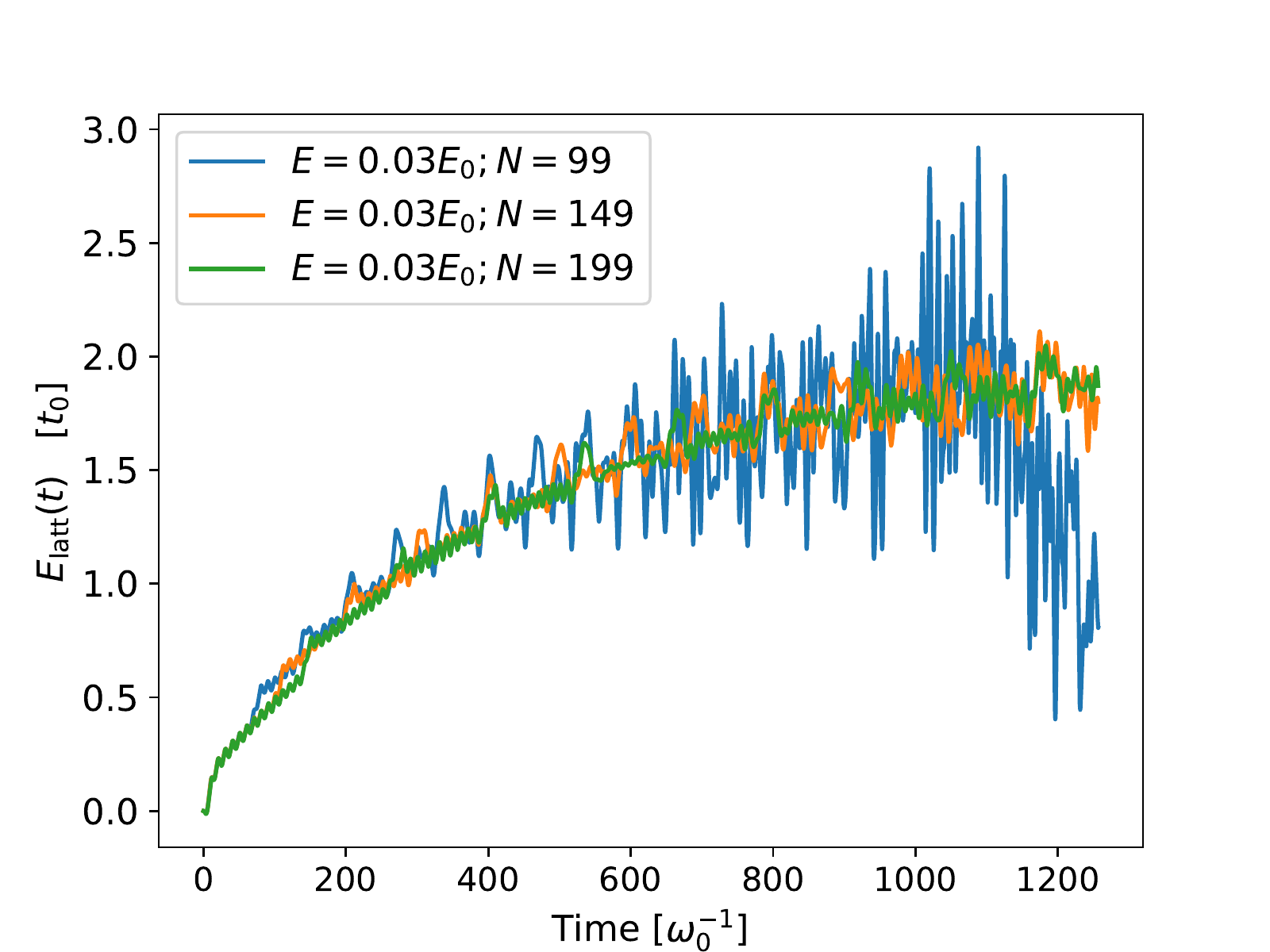}
\caption{Time evolution of : (a) the center of mass $x_{c}\left(t\right)$, (b) the projection $p_{0}\left(t\right)=\left|\left\langle \psi_{0}\left(t\right)|\phi_{0}\left(t\right)\right\rangle \right|^{2}$, and (c) the total potential energy of the lattice for three
different system sizes $N=99,$149 and 199, and for an external field
$E=0.03\ E_{0}$. These plots suggests that an increase of the number of sites $N$ results in an increase of the soliton stability.\label{fig:finite_size_E0.03}}
\end{figure}

\begin{figure}[h]
\includegraphics[scale=0.5]{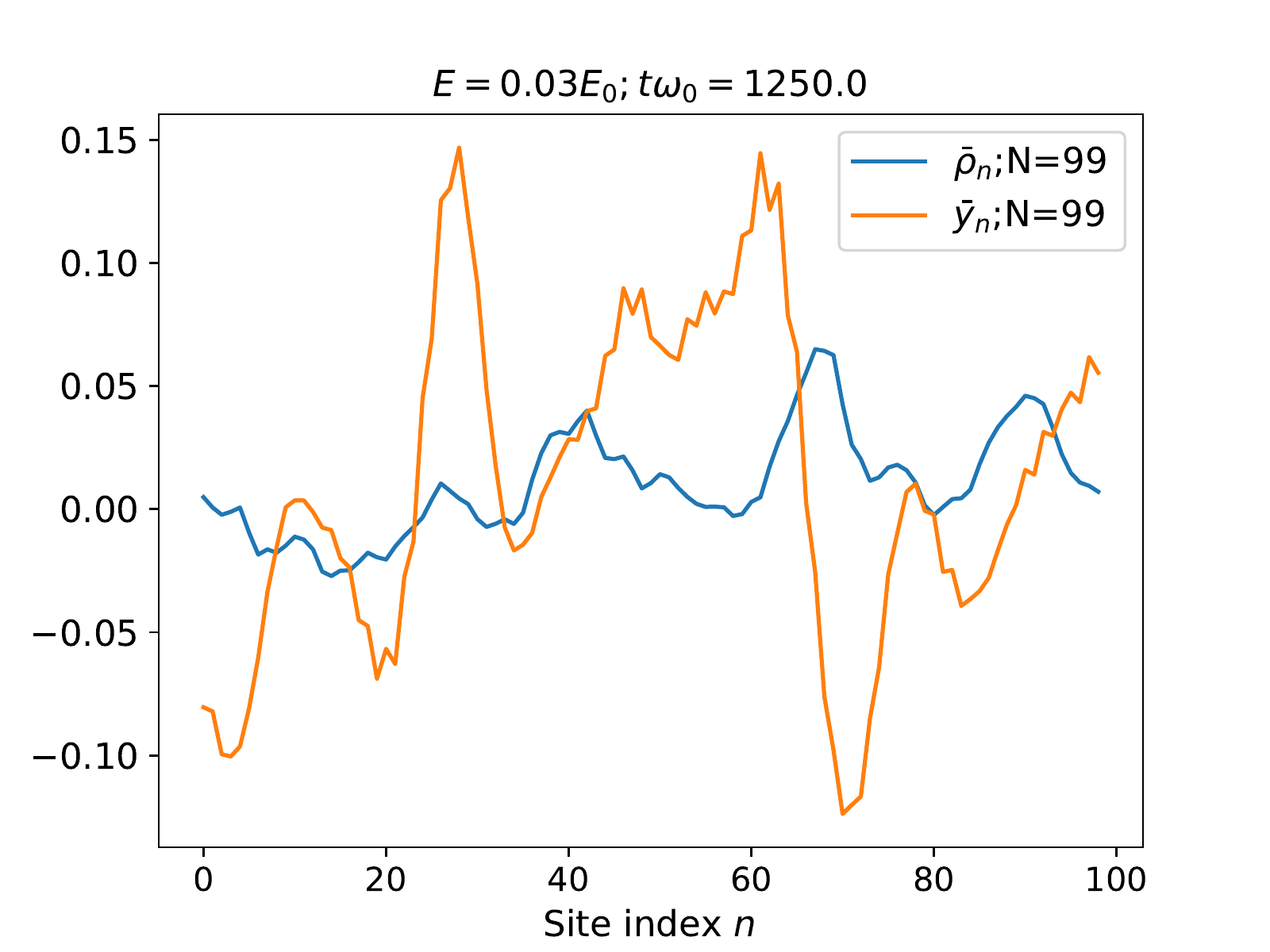}\\
\includegraphics[scale=0.5]{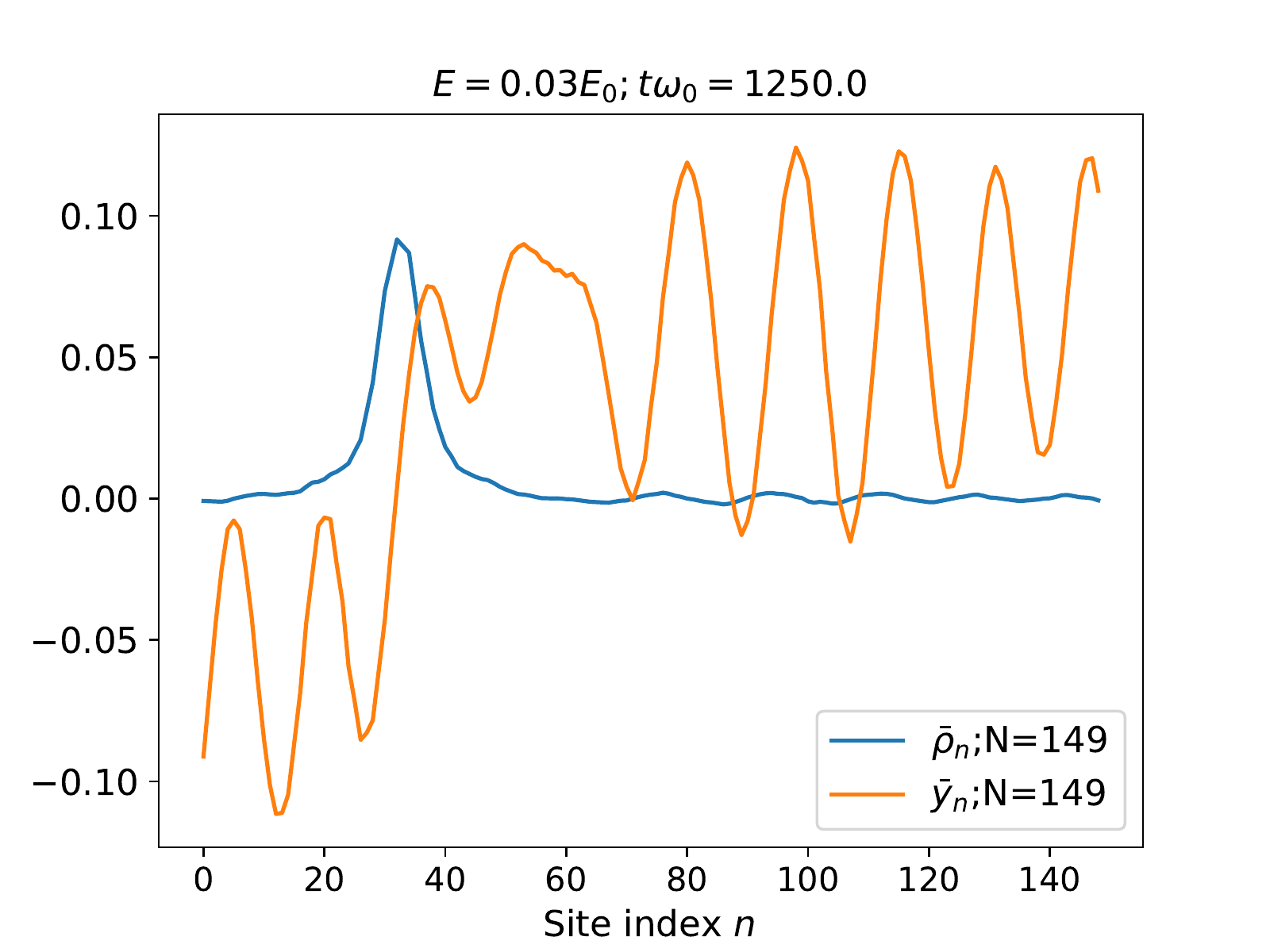}\\
\includegraphics[scale=0.5]{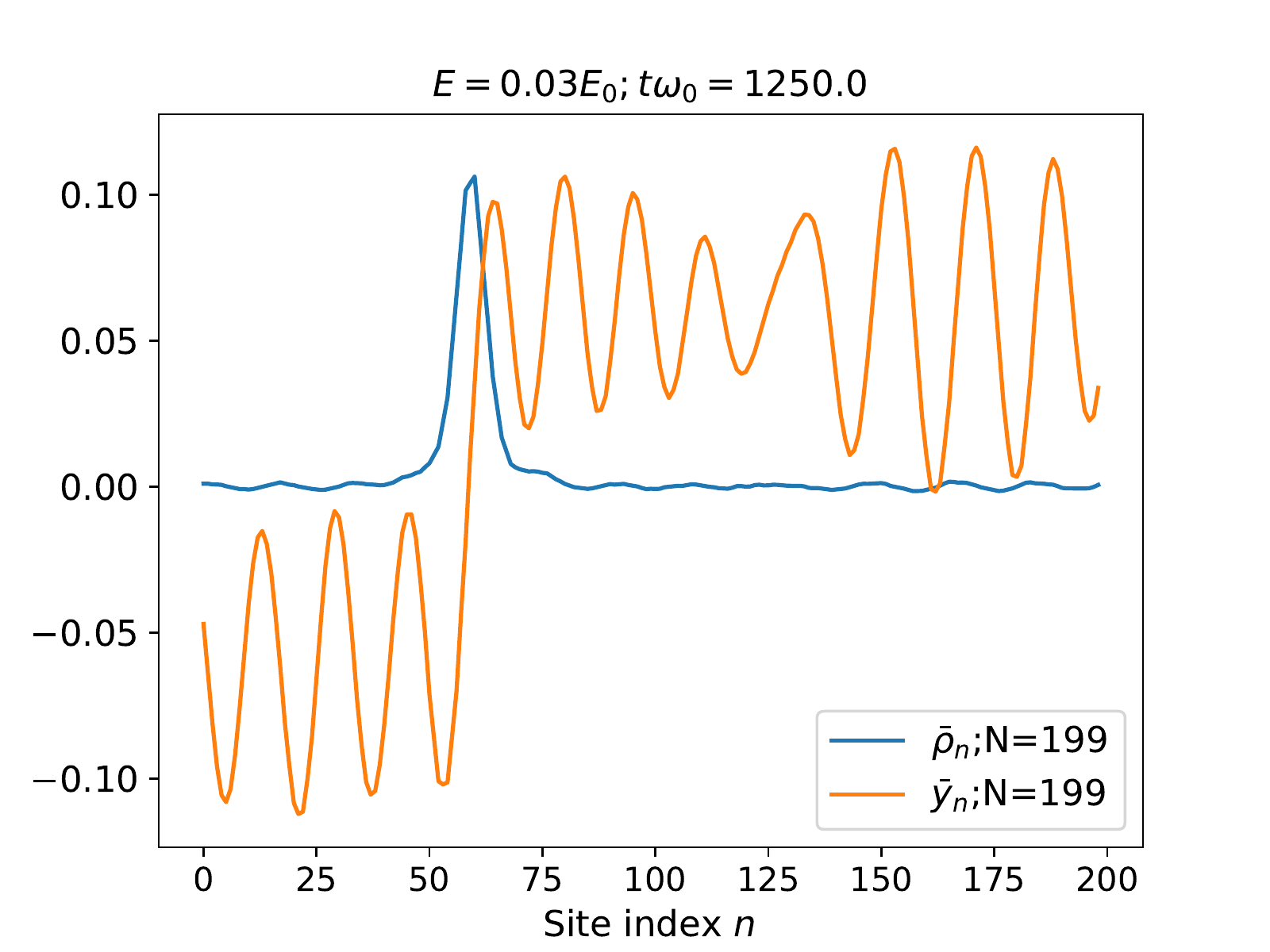}
\caption{Spatial profiles of the excess charge density $\bar{\rho}_{n}\left(t\right)$ and the
lattice deformation profile $\bar{y_{n}}\left(t\right)$ at the final
time step of the simulation for each value of $N$. \label{fig:rho_y_finite_size}}
\end{figure}

To study the dynamics and stability of moving solitons as a function of $N$, we have computed: a) the soliton center-of-mass position $x_{c}\left(t\right)$, b) the projection $p_{0}\left(t\right)=\left|\left\langle \psi_{0}\left(t\right)|\phi_{0}\left(t\right)\right\rangle \right|^{2}$,
c) and the total energy of the lattice $E_\text{latt}(t)=\sum_{n=1}^{N}\left[\frac{M}{2}\dot{u}_n\left(t\right)^2+ \frac{K}{2}y_{n}^{2}\left(t\right) \right]$,
for different system sizes ($N=99$, $N=149$ and $N=199$), and for a fixed value of the external field $E=0.03\ E_0$ (see Fig. \ref{fig:finite_size_E0.03}). In addition, in Fig. \ref{fig:rho_y_finite_size} we show the excess charge density $\bar{\rho}_{n}\left(t\right)$ and lattice deformation profile $\bar{y_{n}}\left(t\right)$ at the final step of the simulation for each of the aforementioned values of $N$. In
all cases we have assumed $N_{\text{el}}=N+1$ (total excess charge
equal to $Q=-e$) in order to compare to our results in previous sections.

These calculations are in qualitative agreement with our previous results with similar parameters. In the case of $x_{c}\left(t\right)$ vs $t$
[see Fig. \ref{fig:finite_size_E0.03}(a)] we even obtain \textit{quantitative} agreement up to the point in which the soliton breaks apart.
However, we note \textit{an increase of the soliton stability as the system
size $N$ increases}. This effect can be rationalized by recalling
that the soliton, being a localized excitation, produces in all cases
a similar perturbation as it travels through the system. In particular,
it transfers the same amount of energy into the lattice as a function
of time, independently of the system size (note the similarity between
the different curves of the total lattice energy $E_\text{latt}(t)$).
However, since the lattice phonons are \textit{delocalized} excitations, the amount of energy
injected by the soliton must be redistributed over chains of different
size in each case, and consequently the amplitude of the deformation
profile $\bar{y}_{n}\left(t\right)$ becomes smaller for larger $N$. This is exactly what  suggests the results shown in Fig. \ref{fig:rho_y_finite_size}. In other words, the amount of energy transferred to the lattice \textit{per site} decreases. 

The results shown in these figures confirm one of the main conclusions
in this work: the mechanism which triggers the instability of the moving
soliton is the large-amplitude fluctuation of the phonon-field, which
eventually makes $\bar{y}_{n}\left(t\right)$ to cross zero and to
generate new electronic states that poisson the (instantaneous) Peierls gap.

Finally, note that a naive theoretical extrapolation of these results would suggest 
that in an infinite system $N\rightarrow\infty$ the soliton would
be always stable (at least in the adiabatic regime of low field $E$). However, this might be 
an artifact of our highly idealized model. Other effects,
such as disorder, finite temperatures, electron-electron interactions,
etc., could change this result and therefore more studies are necessary. In any case, in a real system, the finite length of the polymer should
set the actual value of $N$.

\section{Summary and conclusions}\label{sec:summary}

In this work we have theoretically studied, within the framework of the SSH model, the dynamics of a moving charged soliton in an external electric field, with particular focus on the issue of its stability. While the theoretical study of moving solitons is not new\cite{Su80_Dynamics_of_solitons_in_PA, Mele82_Hot_luminiscence_in_PA, Bishop84_Breathers_in_Polyacetylene, Phillpot87_Dynamics_in_polyyne, ono1990motion,Rakhmanova99_Soliton_dissociation_in_high_electric_field, Johansson04_Nonadiabatic_simulation_of_polaron_dynamics,Takayama80_Continuum_model_for_PA,Guinea84_Dynamics_of_PA_chains, ye1992vibrational}, we stress that many previous works have focused on the stability of solitons moving at a constant velocity $v$ (see, e.g., Refs. \onlinecite{Takayama80_Continuum_model_for_PA, Bishop84_Breathers_in_Polyacetylene,Guinea84_Dynamics_of_PA_chains, Phillpot87_Dynamics_in_polyyne, ye1992vibrational}), a  situation which is conceptually different to ours. To the best of our knowledge a proper definition of an out-of-equilibrium stability criterion has remained an open question.

Our work has been motivated by recent experimental developments where direct evidence of solitonic excitations in tPA molecules deposited on surfaces have been reported\cite{Wang19_Solitons_in_individual_PA_molecules}. These results have triggered renewed interest in this field\cite{HernangomezPerez20_Solitonics_with_PA, cirera2020tailoring, park2022creation}, with promising prospects for the application of solitonic excitations in electronic devices. In this respect, the issue of the stability of a soliton excitation under the effect of an electric field is a relevant question for the design of novel devices based on molecular electronics.		

One of the goals of our work has been to obtain a detailed description of the microscopic mechanisms triggering the dynamical instability of an out-of-equilibrium soliton. While we have focused on the regime of a single soliton (i.e., extremely low-dopant concentration in experimental tPA systems), we mention that a finite concentration of dopants has important additional effects in tPA such as: the presence of a finite density of solitons and formation of a
soliton bands, presence of soliton-(anti)soliton interactions, formation of polarons, metallic behaviour, etc. Due to their enormous complexity, these effects are beyond the scope of our work, and our conclusions cannot be extrapolated to the limit of finite-dopant concentrations.For a comprehensive review of the effects of finite doping, we refer the reader to Ref. \cite{Caldas_2011_Effective_field_theory_of_1D_CH_chains} and references therein.

We have focused on the regime of low external electric field $E$ in which particle-hole excitations are frozen, therefore inducing an adiabatic evolution of the charged solitonic excitation. 
Within the framework of the adiabatic approximation,  the projection onto the instantaneous eigenstates of $H_\text{SSH}$ has enabled us to get useful insight into the microscopic mechanisms triggering the instability during the dynamical evolution of the ``electron + lattice'' system. In particular, we have been able to obtain a ``back-to-back'' comparison between the time-dependent lattice defomation field $y_n(t)$ and the instantaneous energy spectrum of electronic excitations, and to identify the proliferation of soliton-antisoliton pairs, originated in large-amplitude oscillations of the lattice deformation field, as the main destabilizing mechanism. 
 In this context, the projection $p_0(t)=|\langle\psi_{0}(t)|\phi_{0}(t)\rangle|^2$ [Eq. (\ref{eq:p0})], represents a \textit{bona fide} indicator of the stability of a moving soliton. Indeed, the criterion $p_0(t)\simeq 1$ can be interpreted as a ``similarity'' requirement between the evolved zero-energy state $|\psi_0(t)\rangle$ and the instantaneous zero-energy eigenvector of the SSH Hamiltonian $|\phi_{0}(t)\rangle$, which is the ``optimal'' zero-energy state for a given configuration of the deformation field $\bar{y}_n(t)$.  Note that this criterion is generic and valid \emph{beyond} the adiabatic approximation (as long as the evolution of the quantum states is unitary), and can therefore be applied either in equilibrium or in out-of-equilibrium conditions.

From the electronic point of view, the zero-energy state is ``dynamically protected'' by the Peierls gap, and this protection is ensured as far as the instantaneous gap remains stable. Our results suggest (in agreement with previous works\cite{ono1991motion, Kuwabara92_Damping_of_soliton_velocity}) that the energy from the electric field is pumped into the phonon degrees of freedom, and eventually
the oscillations of the deformation field $\bar{y}_n(t)$ acquire enough energy to produce kink-antikink pairs. The proliferation of these kink-antikink pairs ``poisson'' the gap and eventually destroy the aforementioned dynamical protection. This conclusion is consistent with the results obtained in Sec. \ref{sec:finite_size}, where we have found that solitons moving in larger systems are more robust due to a smaller energy \textit{per site} injected into the lattice.

A simplifying approximation used in this work is the mean-field treatment of Eqs. (\ref{eq:newton}), an approximation usually known as the  ``Ehrenfest dynamics''. This is a very common method used also in previous works (see e.g., \cite{Su80_Dynamics_of_solitons_in_PA, Mele82_Hot_luminiscence_in_PA, Bishop84_Breathers_in_Polyacetylene, Phillpot87_Dynamics_in_polyyne, ono1990motion,Rakhmanova99_Soliton_dissociation_in_high_electric_field, Johansson04_Nonadiabatic_simulation_of_polaron_dynamics,Takayama80_Continuum_model_for_PA,Guinea84_Dynamics_of_PA_chains, ye1992vibrational}) to solve the dynamical equations of motion of classical ions coupled to electrons. In particular, although Ehrenfest dynamics can correctly describe the effects of heating of cold electrons by hot ions (a fact that makes it a suitable approximation for computing e.g., electron friction in radiation damage simulations), it is well-known to fail to reproduce the heating of the ions by hot electrons \cite{Horsfield04_Power_dissipation_and_noise}. This asymmetry arises from the absence of quantum fluctuations and the suppression of quantum noise. Therefore, a method that goes beyond the mean-field level would be highly desirable in future works. Different alternative frameworks, such as higher-order expansions of the non-equilibrium Green's functions\cite{stefanucci_van_leeuwen_2013}, or the so-called ``correlated electron-ion dynamics'' (CEID) method \cite{McEniry2010_Correlated_electron_ion_dynamics} are expected to perform better than the Ehrenfest dynamics in this regard. In any case, we expect that the qualitative aspects of our results are not affected  by the limitations of the mean-field approximation, as the microscopic soliton destabilization mechanism is a generic feature, not related to any specific approximation scheme.

As an application, we have revisited previous studies where the dynamics of out-of-equilibrium solitons have been studied\cite{ono1990motion}. In particular, we have studied the evolution of the electronic ``center of mass'' $x_c(t)$ [Eq. \ref{eq:center_of_mass}] and the saturation velocity $v_\text{sat}$ of charged solitons as a function of the electric field $E$. We concluded that the behavior of $x_c(t)$ by itself does not provide reliable information 	about the stability of the soliton excitation (in fact, in some cases it is even misleading, as we have shown in Fig.  \ref{fig:centro_masa_vs_E}), and that it must be complemented by the information provided by $p_0(t)$. In the process of this study, we have obtained new results which are at variance with those of Ref. \onlinecite{ono1990motion}, but which are in agreement with the physical requirement that the saturation velocity of the soliton $v_\text{sat}\rightarrow  0$ in the limit $E\rightarrow 0$.

Our results have been obtained for a regime of electric fields in the range $E\in [0.01,0.1]E_0 $ , which corresponds to $E\in [0.13,1.3]$mV/\AA.  These values are in the regime of operation of optoelectronic devices\citep{Mahani17_Breakdown_of_polarons_in_electric_field}. Therefore our findings may be of interest for experimental and/or technological applications. In addition note that this values of electric fields are well below of the value required for the electrical breakdown of the dielectric [see Eq. (\ref{eq:electric_breakdown})]. This is consistent with figures \ref{fig:espacio_energia_E_p01} and \ref{fig:espacio_energia_E_p1} where we do not see electron-hole excitations (conduction and valence bands do not mix).

Finally, in our study we have crucially assumed the system to be isolated and closed. This simplifying assumption, which is also made in previous works, is clearly unrealistic considering experimental situations. Therefore, the results presented in this work are of interest from the fundamental point of view, but they need to be reconsidered in order to apply them to more realistic systems. In that sense, a very interesting line of research would be to extend these ideas to open systems (e.g., a model of a tPA molecule coupled to a metallic substrate, as in Ref. \onlinecite{Wang19_Solitons_in_individual_PA_molecules}). Additionally, we have assumed zero temperature and the absence of other experimentally important interactions such as disorder, electron-correlation, quantum phonons, etc. These are non-negligible effects that need to be carefully taken into account in more realistic situations (e.g., as in Refs.  \onlinecite{Ogata97_Continuum_model_for_PA_friction, leblanc1984quasirealistic}).

\begin{acknowledgments}
The authors are grateful to Ariel O. Dobry for useful discussions. C.G.S. acknowledges financial support by the Consejo Nacional de Investigaciones Cíentificas y Técnicas (CONICET) through grants PIP 112-2017-0100892CO and PICT-2017-1605. A.M.L. acknowledges financial support from Agencia I+D+i through grant PICT-2017-2081, and from Universidad Nacional de Cuyo through grant SIIP-UNCuyo M083. 
\end{acknowledgments}

\appendix

\section{Critical electric field in adiabatic approximation}
Let $|\psi_\eta(t)\rangle$ be the dynamical single-particle state vector evolved from the initial condition $|\psi_\eta(t=0)\rangle=|\psi_\eta\rangle$, where the $|\psi_\eta\rangle$ is an eigenstate of the Hamiltonian with energy $E_\eta$ [see Eq. \ref{eq:eigenvalue_equation}]
\begin{align}
H_\text{SSH}|\psi_\eta\rangle&=E_\nu |\psi_\eta\rangle,
\end{align}
where we have omitted the spin indices for simplicity. In addition, let $\{|\phi_\nu(t)\rangle \}$ be the set of instantaneous eigenvectors of Hamiltonian of Eq. (\ref{eq:hamiltonian}), which verify the eigenvalue equation
\begin{align}\label{eq:inst_eigenvalue_eq}
H_{\text{SSH}}(t) \left|\phi_{\nu}\left(t\right)\right\rangle &= \epsilon_\nu(t)\left|\phi_{\nu}\left(t\right)\right\rangle,
\end{align}
where $\epsilon_\nu(t)$ are the instantaneous eigenvalues, and where the time appears as a parameter. We assume this basis to be discrete and non-degenerate. Expanding the state vector $|\psi_\eta(t)\rangle$  in the basis $|\phi_\nu(t)\rangle$ yields:
\begin{equation}\label{eq:instantaneous_basis}
|\psi_\eta(t)\rangle = \sum_{\nu}a_{\nu}(t) e^{i\theta_\nu(t)}|\phi_{\nu}(t)\rangle,
\end{equation}
where $\theta_\nu(t)=-\int_0^t \epsilon_\nu(\tau)d\tau/\hbar$ is the dynamical phase factor. Replacing Eq. (\ref{eq:instantaneous_basis}) into the Schr\"odinger equation and contracting with $\langle\phi_\mu(t)|$ yields the equation for the coefficients $a_\mu(t)$:
\begin{equation}\label{eq:schr_eq_coef}
\dot{a}_\mu= -\sum_\nu a_\nu e^{i(\theta_\nu(t)-\theta_\mu(t))}\langle \phi_\mu(t)|\dot{\phi}_\nu (t)\rangle.
\end{equation}
Differentiating Eq. (\ref{eq:inst_eigenvalue_eq}) with respect to time yields,
\begin{equation}
\langle \phi_\mu(t)|\dot{\phi}_\nu(t)\rangle= \frac{\langle\phi_\mu(t)|\dot{H}_\text{SSH}(t)|\phi_\nu(t) \rangle}{\epsilon_\nu(t)-\epsilon_\mu(t)}\qquad \forall \ \mu\neq\nu
\end{equation}

Now, we can rewrite Eq. (\ref{eq:schr_eq_coef}) as 
\begin{align}\label{eq:equation_a}
\dot{a}_\mu(t)&=\sum_{\nu\neq\mu}\frac{\langle\phi_\mu(t)|\dot{H}_\text{SSH}(t)|\phi_\nu (t)\rangle}{\epsilon_\nu(t)-\epsilon_\mu(t)}e^{i(\theta_\nu(t)-\theta_\mu(t))}a_\nu(t),
\end{align}
where we have assmed  $\langle\phi_\mu (t)|\dot{\phi}_\mu (t)\rangle=0$, using the gauge freedom of the instantaneous basis $|\phi_\mu(t)\rangle\rightarrow e^{i\gamma_\mu(t)}|\phi_\mu(t)\rangle$. 	Note that this can always be done, with the exception of Hamiltonians performing a closed loop in parameter space\cite{Berry84}. To a very good approximation, we can assume that this situation never occurs in our system in the regime of parameters of our simulations (indeed, this has been numerically verified). 

The formal solution of Eq. (\ref{eq:equation_a}) can be written in vector form as 
\begin{equation}
\overrightarrow{\mathbf{a}}(t) = T \exp\left\lbrace\int_0^t\overleftrightarrow{\textbf{M}}(\tau)d\tau\right\rbrace \overrightarrow{\mathbf{a}}(0)
\end{equation}
 where $T$ is the time-ordering operator, $\overrightarrow{\mathbf{a}}(t) \equiv (a_1(t), a_2(t), \dots)^T$, and the matrix $\overleftrightarrow{\textbf{M}}(\tau)$ is defined as 
\begin{align}\label{eq:matrixM}
 \left[\overleftrightarrow{\textbf{M}}(\tau)\right]_{\mu\nu}&\equiv\frac{\langle\phi_\mu(\tau)|\dot{H}_\text{SSH}(\tau)|\phi_\nu(\tau)\rangle}{\epsilon_\nu(\tau)-\epsilon_\mu(\tau)}\nonumber \\
 &\times e^{-\frac{i}{\hbar}\int_0^\tau \left(\epsilon_\mu(t^\prime)-\epsilon_\nu(t^\prime) \right)\ dt^\prime}.
\end{align}
The adiabatic theorem Eq. (\ref{eq:adiabatic_theorem}) is recovered in terms of these quantities if 
\begin{equation}
T \exp\left\lbrace\int_0^t\overleftrightarrow{\textbf{M}}(\tau)d\tau\right\rbrace \approx \mathbf{1},
\end{equation}
i.e., the dynamical states are essentially the instantaneous eigenstates of the Hamiltonian. In order for this to occur, the matrix elements in Eq. (\ref{eq:matrixM}) must be very small. More precisely, assuming that the quantities $\epsilon_\mu(t)$ are essentially time-independent under the integral $\int_0^\tau d\tau$ , the integration in Eq. (\ref{eq:matrixM}) can be done very easily, and we obtain the result\citep{messiah}
\begin{equation}
\hbar\frac{\langle\phi_\mu (t)|\dot{H}_\text{SSH}(t)|\phi_\nu(t)\rangle}{({\epsilon_\nu(t) -\epsilon_\mu(t)})^2} \left(2-2\cos\left(\frac{\epsilon_\mu-\epsilon_\nu}{\hbar}t\right)\right)\ll 1,
\end{equation}
which is essentially the adiabatic condition Eq. (\ref{eq:adiabatic_criterion}).

In order to estimate the matrix element $\langle\phi_\mu (t)|\dot{H}_\text{SSH}(t)|\phi_\nu(t)\rangle$ we expand the local site-basis vectors in terms of the instantaneous eigenvector-basis using the formula $c_{n}^{\dagger}=\sum_{\nu}\left[\alpha_{n}^{\nu}\left(t\right)\right]^{*}c_{\nu}^{\dagger}$, where the coefficients $\alpha_{n}^{\nu}\left(t\right)$ are the matrix elements of the unitary change of basis operator $A(t)$. We therefore write
\begin{widetext}
\begin{align*}
\langle\phi_\mu(t)|\dot{H}_\text{SSH}|\phi_\nu(t)\rangle = & \langle 0|c_\mu \biggl\{ \sum_{n}\left[i  t_0 \frac{eaE}{\hbar}-i \alpha y_n(t)  \frac{eaE}{\hbar}-\alpha\dot{y}_n(t)\right] e^{i \frac{eaE}{\hbar}t} c_{n+1}^\dagger c_{n}+ \text{H.c.}\bigg\}  c_\nu^\dagger|0\rangle \\ =&e^{i\frac{eaE}{\hbar}t}\biggl\{ 2it_{0}\frac{eaE}{\hbar}A_{1}^{\mu,\nu}(t) -2i\alpha\frac{eaE}{\hbar} A_{2}^{\mu,\nu}(t)+2\alpha A_{3}^{\mu,\nu}(t) \biggr\}+ \text{H.c.}
\end{align*}
\end{widetext}
where we have defined the quantities
\begin{align*}
A_{1}^{\mu,\nu}(t) \equiv&\sum_{n=1}^{N}\left[\alpha_{n}^{\mu}\left(t\right)\right]^{*}\alpha_{n+1}^{\nu}\left(t\right), \\ A_{2}^{\mu,\nu}(t) \equiv&\sum_{n=1}^{N}y_{n}\left(t\right)\left[\alpha_{n}^{\mu}\left(t\right)\right]^{*}\left[\alpha_{n+1}^{\nu}\left(t\right)\right],\\ 
A_{3}^{\mu,\nu}(t)\equiv &\sum_{n=1}^{N}\left(\frac{dy_{n}\left(t\right)}{dt}\right)\left[\alpha_{n}^{\mu}\left(t\right)\right]^{*}\left[\alpha_{n+1}^{\nu}\left(t\right)\right].
\end{align*}
for which the upper limits: $\left|A_{1}^{\mu,\nu}(t)\right|<1$, $\left|A_{2}^{\mu,\nu}(t)\right|<2u_{0}$ can be imposed. Assuming a quasi static motion of the ions, we can set $d y_n(t)/dt \approx 0$. Then, the following upper limit for the matrix element results
\begin{align*}
\left|\left\langle \phi_{\mu}|\dot{H}_\text{SSH}|\phi_{\nu}\right\rangle \right|<&\sqrt{\left(\frac{4t_{0}eaE}{\hbar}\right)^{2}+\left(\frac{8\alpha u_{0}eaE}{\hbar}\right)^{2}},\\
&= \left(\frac{4t_{0}eaE}{\hbar}\right)\sqrt{1+\left(\frac{2\alpha u_0}{t_0}\right)^2}.
\end{align*}
For the parameters used in this work, the term $\left(\frac{2\alpha u_0}{t_0}\right)^2\approx 0.04$ and can be safely neglected. Finally, particularizing for the zero-energy instantaneous state, the smallest energy difference $\epsilon_\nu(t) -\epsilon_\mu(t)$ corresponds to the Peierls gap $\Delta_g$, and we arrive to the expression
\begin{equation}
\frac{4t_0eaE}{\Delta^2_g} \ll 1,
\end{equation}
which results in the condition Eq. (\ref{eq:stability_condition2})
\begin{equation}
E \ll \frac{\Delta^2_g}{4t_0ea},
\end{equation}

\bibliographystyle{apsrev}

\end{document}